\documentclass[sn-chicago, iicol]{sn-jnl}

\usepackage{amsthm, amsmath, amssymb}
\usepackage{graphicx}
\usepackage{float}
\usepackage{subfigure}
\usepackage{color}
\usepackage{booktabs}
\usepackage{multirow}
\usepackage{geometry}
\usepackage{enumerate}

\jyear{2022}%

\theoremstyle{thmstyleone}%
\newtheorem{theorem}{Theorem}
\newtheorem{lemma}[theorem]{Lemma}
\newtheorem{corollary}[theorem]{Corollary}
\newtheorem{assumption}{Assumption}

\theoremstyle{thmstyletwo}%
\newtheorem{remark}{Remark}%
\theoremstyle{thmstylethree}%

\newcommand{\E}[1]{\mathbb{E}\left[#1\right]}

\newcommand{\bR}{\mathbb{R}}
\newcommand{\di}{\mathrm{d}}
\newcommand{\utau}{\underline{\tau}}
\newcommand{\btau}{\overline{\tau}}
\newcommand{\argmin}{\mathop{\arg\min}}
\newcommand{\abs}[1]{\left\vert#1\right\vert}
\newcommand{\sbr}[1]{\left(#1\right)}
\newcommand{\bbr}[1]{\left\{#1\right\}}
\newcommand{\norm}[1]{\left\lVert #1 \right\rVert }
\newcommand{\sptext}[1]{\; \text{  #1  } \;}
\newcommand{\rbr}[1]{\left[#1\right]}

\newcommand{\D}{\mathcal{D}}
\newcommand{\I}{\mathcal{I}}

\algrenewcommand\algorithmicrequire{\textbf{Input}}
\algrenewcommand\algorithmicensure{\textbf{Output}}

\begin{document}

\title[Renewable CQR Meth. \& Alg.]{Renewable Composite Quantile Method and Algorithm for Nonparametric Models with Streaming Data}


\author[1]{\fnm{Yan} \sur{Chen}}\email{yan\_chen@mail.sdu.edu.cn}

\author[2]{\fnm{Shuixin} \sur{Fang}}\email{fangsx@mail.sdu.edu.cn}

\author*[1]{\fnm{Lu} \sur{Lin}}\email{linlu@sdu.edu.cn, linlu\_sduifs@163.com}

\affil*[1]{\orgdiv{Zhongtai Securities Institute for Financial Studies}, \orgname{Shandong University}, \orgaddress{\street{Shanda Nanlu}, \city{Jinan}, \postcode{250100}, \state{Shandong}, \country{P.R. China}}}

\affil[2]{\orgdiv{School of Mathematics}, \orgname{Shandong University}, \orgaddress{\street{Shanda Nanlu}, \city{Jinan}, \postcode{250100}, \state{Shandong}, \country{P.R. China}}}


\abstract{
We are interested in renewable estimations and algorithms for nonparametric models with streaming data. In our method, the nonparametric function of interest is expressed through a functional depending on a weight function and a conditional distribution function (CDF). The CDF is estimated by renewable kernel estimations combined with function interpolations, based on which we propose the method of renewable weighted composite quantile regression (WCQR). Then we fully use the model structure and obtain new selectors for the weight function, such that the WCQR can achieve asymptotic unbiasness when estimating specific functions in the model. We also propose practical bandwidth selectors for streaming data and find the optimal weight function minimizing the asymptotic variance. The asymptotical results show that our estimator is almost equivalent to the oracle estimator obtained from the entire data together. Besides, our method also enjoys adaptiveness to error distributions, robustness to outliers, and efficiency in both estimation and computation. Simulation studies and real data analyses further confirm our theoretical findings.
}




\maketitle


\section{Introduction}

\subsection{Problem Setup and Challenges}

In this paper, we are interested in nonparametric regression problems for massive data taking the form of streaming data.
Specifically, the considered nonparametric model is that
\begin{equation}\label{eq_model}
    Y = m(X) + \sigma(X)\varepsilon,
\end{equation}
where $Y$ and $X$ are supposed respectively to be scalar response variable and covariate for simplicity; $\varepsilon$ is the random error independent of $X$, and the distribution of $\varepsilon$ is unknown and satisfies $\E{\varepsilon} = 0$ and $\mathrm{Var}[\varepsilon] = 1$; $m: \bR \to \bR$ and $\sigma: \bR \to [0, \infty)$ are both unknown functions.
The considered streaming data consist of a series of cross-sectional data chunks $\D_t = \left\{\sbr{X_{tj}, Y_{tj}}: 1 \leq j \leq n_t\right\}$ for $t = 1, 2, \cdots$,
where all $\sbr{X_{tj}, Y_{tj}}$ are independent and identically distributed (i.i.d.) observations of $(X, Y)$.
In our setting, the data chunks $\D_1, \D_2, \cdots$, are not available simultaneously, but arrive sequentially one after another.

As we know, most conventional statistic algorithms are designed under the premise that the full data can be fitted on the computer memory simultaneously.
However, such a premise is no longer true for streaming data mentioned above.
To deal with streaming data, the online-updating (or renewable) algorithms are widely considered.
For example, at the time $t$, one has obtained a summary statistic $T_t$ of the historical data $\cup_{s \leq t} \D_{s}$.
Then as the new data chunk $\D_{t+1}$ arrives, $T_t$ is updated to $T_{t+1}$ by incremental computation without accessing the historical raw data, i.e., $T_{t+1} = \mathcal{R}\sbr{T_t; \D_{t+1}}$ with $\mathcal{R}$ a function independent with $\cup_{s \leq t} \D_{s}$.
When modified into the above renewable form, the statistics may lose desirable statistical properties, which brings new challenges in designing statistical algorithms for online-updating.

The first challenge arises from the data partitioning.
As we know, nonparametric methods inevitably suffer from estimation bias.
The bias can not be reduced by simply averaging the local estimators from each data chunk, which essentially prevents the renewable estimator from achieving the standard statistical convergence rate.
Hence when designing algorithms for streaming data, it is crucial to sufficiently reduce the estimation bias.

The second challenge lies in the potentially poor quality of steaming data.
Outliers and fat-tailed features are more likely to hide in these massive raw data.
And even worse, it is quite hard to detect or address them, because the relevant procedures usually involve reusing the historical raw data.
Thus it has a significant value for renewable algorithms that the obtained estimator is robust to outliers or adaptive to fat-tailed features.

The third challenge is caused from the exploding data size.
The streaming data source usually generates extremely large amounts of raw data in a short period of time.
To deal with such a rapid data stream, the updating algorithm should be implemented efficiently.

\subsection{Existing Works and Motivations}

There have been many works developed for streaming data.
The existing online-updating methods can be classified into the following categories.
In some restrictive cases, the estimator has a closed-form
expression and the value can be exactly obtained by some recursive updating operations, see, e.g., \cite{Schifano2016, Serhat2009Incremental, Nion2009Adaptive}, etc.
However, it is more often the case that the estimator has no closed-form expression, then iterative algorithms of online-updating are often used to approximate the value of the estimator, see, e.g., \cite{Herbert1951Stochastic, Toulis2014Statistical, Moroshko2015Second, Chen2019Quantile}, etc.
Additionally, several online cumulative frameworks are proposed for likelihood, estimating equations and so on, see, e.g., \cite{Luo2020Renewable, lin2020unified, Wang2022Renewable}, etc.
And there are also some works based on the deep learning techniques, see, \cite{Andri2019Autonomous, Das2019MUSE, Pratama2019Automatic}, to name a few.

In the first scenario, the obtained estimator enjoys exactly the same statistical properties as that of the oracle estimator obtained by using the offline methods together with the full dataset.
However, such a result deeply relies on the closed-form expression of the estimator, which is unavailable for most robust estimators including the quantile estimators.
Without the closed-form expression, the differentiability condition of the objective functions is required to achieve the oracle property, see, e.g., \cite{Luo2020Renewable} and \cite{lin2020unified}.
However, most robust objective functions (e.g., quantile based objective functions) are not differentiable.

In this paper, we will address the above issues by proposing a new method for stream data,
where the renewable algorithm does not rely on the estimator's closed-form expression, but the obtained estimator still achieves the standard statistical convergence rate.
Meanwhile, we also focus on some aspects of aforementioned three challenges and pursue that the proposed method enjoys robustness to outliers or adaptiveness to various error distributions, and the updating algorithm is simple such that it can be implemented efficiently.

\subsection{Contributions and Article Frame}

In this paper, a renewable composite quantile method and algorithm are proposed to estimate the nonparametric functions in the model~\eqref{eq_model} with streaming data.

Inspired by L-Estimation \citep[see, e.g.,][]{Koenker1987Lestimation, Portnoy1989Adaptive, Boente1994Local}, we express the nonparametric function through a functional instead of a closed-form expression.
Here the functional takes the form of an integral depending on a weight function and a conditional distribution function (CDF) of $Y$.
Then the renewable estimation is attained by two steps:
\begin{enumerate}
    \item {\it Numerical Approximation: } The CDF in the functional is approximated by function interpolations.
    Then the nonparametric function can be approximately expressed by a finite number of function values of the CDF.
    \item {\it Statistical Approximation: } The aforementioned function values of CDF are estimated by kernel estimators, which have closed-from expressions and can be exactly obtained through recursive updating algorithms.
\end{enumerate}
By combining the above numerical and statistical approximations, we finally propose our renewable weighted composite quantile regression (WCQR) method for streaming data.

By the renewable WCQR, the functions $m\sbr{\cdot}$ and $\sigma\sbr{\cdot}$ can be estimated by correctly selecting the weight function.
Specifically speaking, we fully use the structure of the model~\eqref{eq_model} and obtain new selection criterions for the weight functions, under which the renewable estimator can estimate $m\sbr{\cdot}$ or $\sigma\sbr{\cdot}$ asymptotically unbiased.
Further, we deduce the asymptotic distributions of the proposed estimators.
Based on this, a practical bandwidth selector is proposed for the online-updating estimator, and the optimal weight function is also obtained by minimizing the asymptotic variance under the constraint of the above selection criterions.
Finally, our theoretical findings are demonstrated by simulation studies and real data analyses.

Compared with the competitors, our method has the following main virtues:
\begin{itemize}
    \item[1)] {\it Oracle comparability}.
    Through numerical approximations, our WCQR estimator is assembled from some renewable statistics exactly obtained via online-updating.
    Thanks to this, not only the algorithm gets rid of any restriction on the chunk size or chunk number of the streaming data, but also the obtained estimator enjoys almost the same asymptotic properties as that of the oracle estimator obtained on the full data set.
    \item[2)] {\it Robustness}.
    Benefit from robust feature of the quantile estimation, when the model has a symmetric error distribution, our regression method enjoys robustness compared with the common methods such as ordinary least squares.
    \item[3)] {\it Model adaptiveness}.
    With model-based weight functions, our estimation method is adaptive to symmetric or asymmetric models.
    Different from existing methods, the selection criterions of the weight functions are directly established on the structure of the nonparametric models instead of the information of the errors.
    Thus in our method, the weight selection does not rely on any pilot estimations for the error distributions and the model adaptiveness can be preserved in case of streaming data.
    \item[4)] {\it Estimation efficiency}.
    Under the above weight criterions, we find the optimal weight function, under which our renewable estimator enjoys minimized asymptotic variance in estimating $m\sbr{x}$ and $\sigma\sbr{x}$.
    \item[5)] {\it Computational efficiency}.
    Thanks to the closed-form expression of the kernel estimators, our algorithm is quite simple in updating procedures with no need for solving any optimization problems or nonlinear equations. This feature is particularly desirable for rapid data stream.
\end{itemize}

The paper is then organized in the following way.
In Section~\ref{sec_prel}, we give some preliminaries above the existing composite quantile estimations and the L-Estimation.
In Section~\ref{sec_meth}, the main idea of our methodologies is introduced in detail, including the computation of the renewable WCQR estimator, the selection of the weight functions, and some specific estimators and detailed renewable algorithms for estimating $m\sbr{x}$ and $\sigma\sbr{x}$.
In Section~\ref{sec_theo}, the asymptotic properties of the proposed method are established; the selector of online-updating bandwidth is proposed and the optimal selection of the weight function is discussed.
Section~\ref{sec_simu} contains comprehensive simulation studies and real data analyses to further demonstrate the desirable performance of the proposed estimators and algorithms.
Some algorithms, lemmas, and technical proofs are deferred to the supplementary material.

\subsection{Notations}

For a random variable $Z$, denote by $f_Z(\cdot)$ and $F_Z(\cdot)$ the the probability density function (PDF) and the distribution function of $Z$, respectively.
Denote by $Q_{Y\vert x}(\tau)$ the conditional $100\tau\%$ quantile of $Y$ given $X=x \in \bR$ and $\tau \in (0, 1)$, i.e. $Q_{Y\vert x}(\tau) = \inf\left\{t: \mathbb{P}\sbr{Y \leq t\vert X=x} \geq \tau \right\}$.
Denote by $F_{Y\vert x}\sbr{\cdot}$ the conditional distribution function (CDF) of $Y$ given $X = x$, i.e., $F_{Y\vert x}\sbr{y} = \mathbb{P}\sbr{Y<y\vert X=x}$.
Denote by $f_{Y\vert x}\sbr{\cdot}$ the conditional PDF of $Y$ given $X = x$.
For $a, b \in \bR$, denote $a \land b = \min\bbr{a, b}$.
For indexes $i$ and $j$, denote by $\delta_{ij}$ the kronecker symbol, i.e., $\delta_{ij} = 1$ if $i = j$ and $\delta_{ij} = 0$ if $i \neq j$.
Denote by $\mathrm{I}_{\mathrm{d}}\sbr{\cdot}$ the identity function, i.e., $\mathrm{I}_{\mathrm{d}}\sbr{y} = y$.
Denote by $I\sbr{\cdot}$ the indicative function, i.e., if the proposition $\mathcal{P}$ is true, $I\sbr{\mathcal{P}} = 1$, and otherwise, $I\sbr{\mathcal{P}} = 0$.
For a function $f: \bR \to \bR$, denote by $\mathrm{Supp} \bbr{f\sbr{\cdot}}$ the support set of $f$; for $I_0 \subset \bR$ and $1 \leq p < \infty$, denote $\norm{f}_{p, I_0} = (\int_{I_0}\abs{f(y)}^p \di y)^{1/p}$ and $\norm{f}_{\infty, I_0} = \sup_{y \in I_0} \abs{f(y)}$.
For a finite point set $G \subset \bR$, denote $\Delta\sbr{G} = \max_{y_i \in G} \; (\min_{y_j \in G \backslash \bbr{y_i}} \abs{y_i - y_j})$, i.e., the maximum spacing between  any two adjacent points in $G$;
denote by $\#G$ the number of elements in $G$, and denote by $\min G$ (resp. $\max G$) the minimum  (resp. maximum) of $G$.
For two sets $G_1$ and $G_2$, denote $G_1 \backslash G_2 = \bbr{y: y \in G_1, y \notin G_2}$.
For a set $E \subset \bR$ and a function $f: \bR \to \bR$, denote $f\sbr{E} = \bbr{f(x): x \in E}$; denote $I\sbr{E} = [\inf E, \sup E]$ the closed interval generated by $E$.

\section{Preliminary}\label{sec_prel}

In this section, we will briefly review some fundamental works, from which we obtain inspirations for our method.

\subsection{Composite Quantile Regression and L-Estimation}\label{sec_CQR_Lest}

To estimate the regression function $m(x)$, the well-known composite quantile regression (CQR) estimator \citep[see, e.g.,][]{Zou2008Composite, Kai2010} takes the form of
\begin{equation}
    \widehat{m}_{\mathrm{cqr}}(x) = \frac{1}{q} \sum_{i=1}^q \widehat{Q}_{Y\vert x}\sbr{\tau_i},
\end{equation}
where $\widehat{Q}_{Y\vert x}\sbr{\tau_i}$ are some consistent estimators of the quantile $Q_{Y\vert x}\sbr{\tau_i}$, and $\tau_i$ are quantile levels chosen as $\tau_i = i/(q+1)$.
When the PDF of $\varepsilon$ is symmetric around $0$, one can expect $m\sbr{x} = 1/q \sum_{i=1}^q Q_{Y\vert x}\sbr{\tau_i}$ implying that $\widehat{m}_{\mathrm{cqr}}(x)$ is an asymptotic unbiased estimator of $m\sbr{x}$.
However, when the error is general, the aforementioned equality is not necessarily true and the naive CQR estimator may suffer from non-negligible bias.
To address this issue, \cite{Sun2013Weighted} extended the CQR estimator to the WCQR estimator in from of
\begin{equation}\label{eq_mwcqr}
    \widehat{m}_{\mathrm{wcqr}}(x) = \sum_{i=1}^q \omega_i \widehat{Q}_{Y\vert x}\sbr{\tau_i},
\end{equation}
where $\omega_i$ are weights selected to satisfy $\sum_{i=1}^q \omega_i = 1$ and $\sum_{i=1}^q \omega_i F_{\varepsilon}^{-1}\sbr{\tau_i} = 0$, such that the equality $m\sbr{x} = \sum_{i=1}^q \omega_i Q_{Y\vert x}\sbr{\tau_i}$ is established and meanwhile, the non-negligible bias of $\widehat{m}_{\mathrm{wcqr}}(x)$ is eliminated.

The CQR and WCQR estimators can be expressed as L-estimators \citep[see, e.g.,][]{Gutenbrunner1992Regression, Koenker1994Lesti} taking the form
\begin{equation}\label{eq_mLxnu}
    \widehat{m}_{\mathrm{L}}(x; \nu) = \int_{[0, 1]} \widehat{Q}_{Y\vert x} \sbr{\tau} \di \nu\sbr{\tau},
\end{equation}
with $\nu$ a measure on $[0, 1]$.
To cover the CQR (resp. WCQR) estimators, the measure of L-estimators can be particularly chosen as $\nu = 1/q \sum_{i=1}^q \delta_{\tau_i}$ (resp. $\nu = \sum_{i=1}^q \omega_i \delta_{\tau_i}$) with $\delta_{\tau_i}$ a unit mass on $\tau_i$.
Additionally, when the measure $\nu$ is well chosen based on the error distribution in \eqref{eq_model}, the L-estimator can efficiently (or robustly) estimate $m\sbr{x}$ and $\sigma\sbr{x}$ \citep[see, e.g.,][etc]{Serfling1980Approximation, Koenker1987Lestimation, Portnoy1989Adaptive, Koenker2005}.

In this paper, we still call $\widehat{m}_{\mathrm{L}}(x; \nu)$ the WCQR estimator to highlight the weighted composite of quantile estimators.

\subsection{Local Polynomial Interpolation}\label{sec_LPI}

We will briefly review the main idea of local polynomial interpolation (LPI), which is a fundamental tool for numerical approximations and will be applied in our method.
For more details about the interpolation techniques and their applications to stochastic computing, readers may refer to \cite{gautschi2011numerical, sauer2011numerical, burden2015numerical, Zhao2006new, Zhao2014New, Fu2017Efficient}, etc.

Suppose that $\sbr{y_1, f_1}, \cdots, \sbr{y_q, f_q}$ are given data point generated from an unknown objective function $f: \bR \to \bR$, i.e., $f_i = f\sbr{y_i}$ for $i = 1, \cdots, q$.
Denote the node set $G_x = \bbr{y_i}_{i=1}^q$ and the elements $y_i$ are called the interpolation nodes.

The so-called LPI aims at finding a piecewise polynomial function passing through all the given data points of $f\sbr{\cdot}$.
To achieve this, for $1 \leq l \leq q$ and $y \in \bR$, we denote by $\mathcal{N}_l \sbr{y, G_x}$ the set consisting of the $l$ nearest interpolation nodes around $y$, or strictly speaking, $\mathcal{N}_l \sbr{y, G_x} = G_0$ with $G_0$ the unique set satisfying:
\begin{enumerate}
    \item[a)] $\# G_0 = l$, $G_0 \subset G_x$;
    \item[b)] for any $y^{\prime} \in G_0$ and $y^{\prime\prime} \in G_x\backslash G_0$, it holds that $\abs{y - y^{\prime}} \leq \abs{y - y^{\prime \prime}}$, and whenever the equality holds, $y^{\prime} < y^{\prime\prime}$.
\end{enumerate}
Then we define the $l$th-degree LPI basis functions as
\begin{equation}\label{eq_Liy}
    L\sbr{y, y_i; l, G_x} = \ell_i\sbr{y} I\sbr{y_i \in \mathcal{N}_{l+1}\sbr{y, G_x}}
\end{equation}
for $y_i \in G_x$ with $\ell_i\sbr{\cdot}$ the $l$th-degree Lagrange interpolating polynomials defined as
\begin{equation*}
    \ell_i\sbr{y} = \frac{\prod_{y_j \in \mathcal{N}_{l+1}\sbr{y, G_x}\backslash \bbr{y_i}} \sbr{y - y_j}}{\prod_{y_j \in \mathcal{N}_{l+1}\sbr{y, G_x}\backslash \bbr{y_i}} \sbr{y_i - y_j}}.
\end{equation*}
Using the basis functions, the $l$th-degree LPI function  is constructed as
\begin{equation}\label{eq_lpiPy}
    \I_{l} f\sbr{y} = \sum_{y_i \in G_x} f\sbr{y_i } L\sbr{y, y_i; l, G_x}.
\end{equation}
Here and in the following context, we formally use $\I_l$ to denote the $l$th-degree LPI operator in the sense that $\I_l f\sbr{\cdot}$ is a $l$th-degree LPI function obtained by \eqref{eq_lpiPy}.

It is easy to verify that $L\sbr{y_i, y_j; l, G_x} = \delta_{ij}$, implying that $\I_l f\sbr{\cdot}$ indeed passes through the points $\bbr{\sbr{y_i, f_i}}_{i=1}^q$.
The accuracy of the LPI function approximating the objective function can be guaranteed by the following standard result (see, e.g., Theorem 3.3 of \citealt{burden2015numerical} or Theorem 3.3 of \citealt{sauer2011numerical}):
\begin{lemma}[\citealt{burden2015numerical, sauer2011numerical}]\label{lemm_LPI}
    Suppose that $f\sbr{\cdot}$ has continuous derivatives upto order $l+1$, and $\I_l f(\cdot)$ is given in \eqref{eq_lpiPy}.
    Then it holds that
    \begin{equation*}
        f(y) = \I_l f(y) + \frac{f^{(l+1)}(c)}{\sbr{l+1}!}\prod_{y_i \in \mathcal{N}_{l+1}\sbr{y, G_x}} \sbr{y - y_i},
    \end{equation*}
    where $c$ is a number depends on $y$ and lies in $\rbr{\min \overline{\mathcal{N}}_{l+1}\sbr{y}, \max \overline{\mathcal{N}}_{l+1}\sbr{y}}$  with $\overline{\mathcal{N}}_{l+1}\sbr{y} = \mathcal{N}_{l+1}\sbr{y, G_x} \cup \bbr{y}$.
\end{lemma}

\section{Methodology}\label{sec_meth}

In this section, we aim at proposing a renewable WCQR estimation to estimate $m\sbr{x}$ and $\sigma\sbr{x}$ for $x \in I_*$ with $I_*$ a bounded interval on $\bR$.

\subsection{Renewable WCQR Estimation}

We start form a special case of \eqref{eq_mLxnu}, where the measure $\nu$ exists a density function $J\sbr{\cdot}$ on $[0, 1]$, and the WCQR estimator $\widehat{m}_{\mathrm{L}}(x; \nu)$ can be expressed as
\begin{equation}\label{eq_hatmx}
    \widehat{r}(x; J) = \int_{[0, 1]} J\sbr{\tau}\widehat{Q}_{Y\vert x} \sbr{\tau} \di \tau.
\end{equation}
By selecting appropriate $J\sbr{\cdot}$, the estimator $\widehat{r}(x; J)$ is able to estimate a kind of parameters that can expressed as
\begin{equation}\label{eq_theta}
    r(x; J) = \int_{[0, 1]} J\sbr{\tau} Q_{Y\vert x} \sbr{\tau} \di \tau.
\end{equation}

To obtain $\widehat{r}(x; J)$, the conventional approach is to minimize an $L_1$-norm loss function characterized by the check function $\rho_{\tau}(u) = u\sbr{\tau - I\sbr{u \leq 0}}$.
However, solving such a non-smooth minimization problem is quite difficult when the data are of the form of streaming data sets.
To address this issue, we manage to avoid estimating the quantiles in \eqref{eq_mwcqr}.
Thus we introduce the substitution: $y = \widehat{Q}_{Y\vert x}\sbr{\tau}$, i.e., $\tau = \widehat{F}_{Y\vert x}\sbr{y}$ and rewrite~\eqref{eq_hatmx} into
\begin{equation}\label{eq_defmJ}
    \widehat{r}(x; J) = \int_{\bR}  y J\sbr{\widehat{F}_{Y\vert x}\sbr{y}} \di \widehat{F}_{Y\vert x}\sbr{y},
\end{equation}
where $\widehat{F}_{Y\vert x}\sbr{y}$ is an estimator of the CDF $F_{Y\vert x}\sbr{y}$.

Now the the key problem is to obtain a renewable estimation of $F_{Y\vert x}\sbr{\cdot}$.
To this end, we approximate $F_{Y\vert x}\sbr{\cdot}$ by its LPI function given by
\begin{equation}\label{eq_intpFYx}
    \I_l F_{Y\vert x}\sbr{y} = \sum_{y_i \in G_x} F_{Y \vert x}\sbr{y_i} L\sbr{y, y_i; l, G_x},
\end{equation}
where $G_x = \bbr{y_i}_{i=1}^q$ is the set of interpolation nodes chosen for $x \in I_*$ and $L\sbr{\cdot, y_i; l, G_x}$ are PLI basis functions defined in \eqref{eq_Liy}.
Then the unknown values $F_{Y \vert x}\sbr{y_i}$ can be estimated by their empirical
analogues obtained from the streaming data sets $\D_1, \D_2, \cdots$, i.e.,
\begin{equation}\label{eq_hatFYx}
    \widehat{F}_{Y \vert x, t}\sbr{y_i} = \frac{\widehat{S}_{Y \vert x, t} \sbr{y_i}}{\widehat{f}_{X, t} \sbr{x}} \sptext{for} y_i \in G_x,
\end{equation}
where $\widehat{f}_{X, t} \sbr{x}$ and $\widehat{S}_{Y \vert x, t} \sbr{y_i}$ are renewable statistics obtained by
\begin{align}
    N_t =\;& N_{t-1} + n_t, \notag \\
    \widehat{f}_{X, t} \sbr{x} =\;& \frac{N_{t-1}}{N_t}\widehat{f}_{X, t-1}\sbr{x} \notag \\
    &+ \frac{1}{N_t}\sum_{j=1}^{n_t} K_{h_t}\left(X_{tj} - x\right), \label{eq_St}\\
    \widehat{S}_{Y\vert x, t} \sbr{y_i} =\;& \frac{N_{t-1}}{N_t} \widehat{S}_{Y\vert x, t-1} \sbr{y_i} \notag \\
    + \frac{1}{N_t}& \sum_{j=1}^{n_t} I\sbr{Y_{tj} < y_i} K_{h_t}\left(X_{tj} - x\right) \label{eq_Ct}
\end{align}
with initial values $N_0 = \widehat{f}_{X, 0} \sbr{x} = \widehat{S}_{Y\vert x, 0} \sbr{y_i} = 0$ and the bandwidths $h_t > 0$.
Based on \eqref{eq_intpFYx}, we plug in the estimators $\widehat{F}_{Y \vert x, t}\sbr{y_i}$ of $F_{Y \vert x}\sbr{y_i}$ and obtain the interpolated empirical CDF as follows
\begin{equation}\label{eq_hatintpFYx}
    \I_l \widehat{F}_{Y\vert x, t}\sbr{y} = \sum_{y_i \in G_x} \widehat{F}_{Y \vert x, t}\sbr{y_i} L\sbr{y, y_i; l, G_x}.
\end{equation}

By \eqref{eq_defmJ} with $\I_l \widehat{F}_{Y\vert x, t}(y)$ in place of $\widehat{F}_{Y\vert x}\sbr{y}$, we can obtain the renewable WCQR estimator as
\begin{equation}\label{eq_widehatthetaF}
    \widetilde{r}_t(x; J) = \int_{\bR}  y J\sbr{\I_l \widehat{F}_{Y\vert x, t}\sbr{y}} \di \I_l \widehat{F}_{Y\vert x, t}\sbr{y}.
\end{equation}
Since the expression of $\I_l \widehat{F}_{Y\vert x, t}\sbr{\cdot}$ is known, the integral in \eqref{eq_widehatthetaF} can be accurately approximated by numerical integrations,
e.g., the well-known trapezoidal rule, Simpson rule and Romberg integration, etc (see, e.g., Section 3 of \citealt{gautschi2011numerical}).

By applying LPI on the $x$-axis, we can extend the pointwise estimator $\widetilde{r}_t(x; J)$ to estimate the function $r(\cdot\;; J)$ on the entire interval $x \in I_*$.
Specifically, we can approximate $r(\cdot\;; J)$ by its LPI function, i.e.,
\begin{equation}\label{eq_Gstar}
    \I_l r(x; J) = \sum_{x_i \in G_*} r(x_i; J) L\sbr{x, x_i; l, G_*},
\end{equation}
where $G_* = \bbr{x_i}_{i=1}^{\bar{q}}$ is a set of grid points introduced on the interval $I_*$, and typically $x_i$ can be chosen as equal-spaced grid points on $I_*$.
With $r(x_i; J)$ estimated by $\widetilde{r}_t (x_i; J)$, we can obtain the renewable interpolated WCQR estimator as
\begin{equation}\label{eq_Iltilder}
    \I_l \widetilde{r}_t (x; J) = \sum_{x_i \in G} \widetilde{r}_t(x_i; J) L\sbr{x, x_i; l, G_*}
\end{equation}
for $x \in I_*$.
In Section~\ref{sec_specwcqr}, we will see that $\I_l \widetilde{r}_t (\cdot\;; J)$ plays a role in selecting the weight function $J\sbr{\cdot}$ for streaming data.

For the restriction on the weight function in \eqref{eq_widehatthetaF}, we have the following remark.
\begin{remark}\label{rmk_ResWeight}
    By Lemma~\ref{lemm_LPI}, we can conclude that
    \begin{equation*}
        \begin{aligned}
            & \|\I_l F_{Y \vert x} - F_{Y \vert x}\|_{\infty, I\sbr{G_x}} \\
            \leq\;& \| F_{Y \vert x}^{\sbr{l+1}}\|_{\infty, I\sbr{G_x}}\abs{\Delta\sbr{G_x}}^{l+1}
        \end{aligned}
    \end{equation*}
    with $I\sbr{G_x} = [\min G_x, \max G_x]$.
    Thus if $F_{Y\vert x}(\cdot)$ is sufficiently smooth and the nodes are dense enough, the error caused from LPI can be negligible on $I\sbr{G_x}$.
    However, when $y$ moves away from the interval $I\sbr{G_x}$, the LPI approximation cannot guarantee its accuracy.
    Fortunately, we can select appropriate $J\sbr{\cdot}$ to suppress the error of LPI when $y$ lies outside $I\sbr{G_x}$.
    Specifically, for the renewable WCQR estimator $\widetilde{r}_t(x; J)$, we can select $J\sbr{\cdot}$ satisfying
    \begin{equation}\label{eq_suppJsubset}
        \mathrm{Supp} \bbr{J\sbr{\cdot}} \subset \I_l \widehat{F}_{Y\vert x, t}^{-1}\sbr{I\sbr{G_x}}.
    \end{equation}
    And for the interpolated WCQR estimator $\I_l \widetilde{r}_t (\cdot\;; J)$, we can select $J\sbr{\cdot}$ satisfying \eqref{eq_suppJsubset} for all $x \in G_*$ with $G_*$ satisfying $I\sbr{G_*} \supset I_*$.
    Given $I\sbr{G_x}$ wide enough, the above restrictions are mild in robust estimations,
    because $\mathrm{Supp}\bbr{J\sbr{\cdot}} \subset (0, 1)$ is a natural condition to guarantee the robustness of $\widehat{r}\sbr{x; J}$ \citep[see, e.g., Section~8.1.3 of][]{Serfling1980Approximation}.
\end{remark}

In the remainder of this section, we mainly discuss the selection of the weight function $J(\cdot)$, which plays a key role in reducing the estimation bias and variance of our renewable WCQR estimator.

\subsection{Model-Based Weight Selections}\label{sec_modelweighsel}

To estimate $m\sbr{x}$ (resp. $\sigma\sbr{x}$) by renewable WCQR estimation, we should select appropriate weight functions $J\sbr{\cdot}$ to establish the equality $r\sbr{x; J} = m\sbr{x}$ (resp. $\sigma\sbr{x}$).
To this end, recalling \eqref{eq_theta} and using the relation $Q_{Y\vert x} \sbr{\tau} = m(x) + \sigma(x) F_{\varepsilon}^{-1} \sbr{\tau}$ obtained from the model~\eqref{eq_model}, we can deduce
\begin{equation}\label{dec_rxJ}
    \begin{aligned}
        r(x; J) =\;& m(x) \int_{[0, 1]} J(\tau) \di \tau \\
        &+ \sigma(x) \int_{[0, 1]} J(\tau) F_{\varepsilon}^{-1}\sbr{\tau} \di \tau,
    \end{aligned}
\end{equation}
which yields the conditions for estimating $m\sbr{x}$:
\begin{equation}\label{eq_cons_J}
    \begin{aligned}
        \mathrm{C}_{m1}: &\int_{[0, 1]} J(\tau) \di \tau = 1, \\
        \mathrm{C}_{m2}: &\int_{[0, 1]} J(\tau) F_{\varepsilon}^{-1}\sbr{\tau} \di \tau = 0,
    \end{aligned}
\end{equation}
and the conditions for estimating $\sigma\sbr{x}$:
\begin{equation}\label{eq_cons_Jsgm}
    \begin{aligned}
        \mathrm{C}_{\sigma 1}: &\int_{[0, 1]} J(\tau) \di \tau = 0, \\
        \mathrm{C}_{\sigma 2}: &\int_{[0, 1]} J(\tau) F_{\varepsilon}^{-1}\sbr{\tau} \di \tau = 1.
    \end{aligned}
\end{equation}

Among the above conditions, $\mathrm{C}_{m1}$ and $\mathrm{C}_{\sigma 1}$ can be easily satisfied.
However, $\mathrm{C}_{m2}$ and $\mathrm{C}_{\sigma 2}$ are related to the unknown quantile function $F_{\varepsilon}^{-1}(\cdot)$.
Unless the error distribution is known to be symmetric, it is quite difficult to choose $J(\cdot)$ in an renewable manner; see the following remark.

\begin{remark}\label{rmk_pseudoJ}
    In the existing works, e.g., \cite{Sun2013Weighted, LinComposite2019, Jiang2016Single}, the condition $\mathrm{C}_{m2}$ is fulfilled empirically by replacing the unknown function $F_{\varepsilon}^{-1}(\cdot)$ with its estimator $\widehat{F}_{\varepsilon}^{-1}(\cdot)$.
    Here $\widehat{F}_{\varepsilon}^{-1}(\cdot)$ is the sample quantile function of the ``pseudo'' samples $\widehat{\varepsilon}_{ti} = \sbr{Y_{ti} - \widehat{m}\sbr{X_{ti}}}/\widehat{\sigma}\sbr{X_{ti}}$, where $\widehat{m}\sbr{\cdot}$ and $\widehat{\sigma}\sbr{\cdot}$ are some pilot estimators of $m\sbr{\cdot}$ and $\sigma \sbr{\cdot}$.
    The generation of pseudo samples involves reusing the historical raw data and requires sophisticated computations,
    which are hardly to be implemented for streaming data.
\end{remark}

To avoid the problem mentioned above, we fully use the structure of the Model~\eqref{eq_model} and obtain the following important lemma, which gives an alternative way to fulfill the conditions $\mathrm{C}_{m2}$ and $\mathrm{C}_{\sigma 2}$.

\begin{lemma}\label{lemm_intJF0}
    Let $W: \bR \to \bR$ be a function satisfying $\E{W\sbr{X} \sigma \sbr{X}} > 0$, then the following results hold:
    \begin{enumerate}
        \item[i)] if $\mathrm{C}_{m1}$ holds, then
            \begin{equation*}
                \begin{aligned}
                    &\E{W\sbr{X} \sigma \sbr{X}} \int_{[0, 1]} J(\tau) F_{\varepsilon}^{-1} \sbr{\tau} \di \tau \\
                    =\;& \E{W(X) \sbr{ r\sbr{X; J} - Y}};
                \end{aligned}
            \end{equation*}
        \item[ii)] if $\mathrm{C}_{\sigma 1}$ holds, then
            \begin{equation*}
                \begin{aligned}
                    & \E{W(X) r^2\sbr{X; J}} \sbr{\int_{[0, 1]} J(\tau) F_{\varepsilon}^{-1} \sbr{\tau} \di \tau}^2 \\
                    =\;& \E{W\sbr{X} \sbr{Y^2 - m^2 \sbr{X}}}.
                \end{aligned}
            \end{equation*}
    \end{enumerate}
\end{lemma}
Lemma~\ref{lemm_intJF0} suggests the following alternative conditions on the weight function:
\begin{align}
    \mathrm{C}_{m 2}^{\prime}:&\; \E{W(X) r\sbr{X; J}} = \E{W(X)Y}, \label{eq_C2p} \\
    \mathrm{C}_{\sigma 2}^{\prime}:&\; \E{W(X) r^2\sbr{X; J}} \notag \\
    & = \E{W(X) \sbr{Y^2 - m^2 \sbr{X}}} \label{eq_Cs2p}.
\end{align}
Lemma~\ref{lemm_intJF0} also shows that $\sbr{\mathrm{C}_{m1}, \mathrm{C}_{m2}} \Leftrightarrow \sbr{\mathrm{C}_{m1}, \mathrm{C}_{m2}^{\prime}}$ and $\sbr{\mathrm{C}_{\sigma 1}, \mathrm{C}_{\sigma 2}} \Leftrightarrow \sbr{\mathrm{C}_{\sigma 1}, \mathrm{C}_{\sigma 2}^{\prime}}$, i.e., the condition $\mathrm{C}_{m2}$ (resp. $\mathrm{C}_{\sigma 2}$) can be equivalently replaced with $\mathrm{C}_{m2}^{\prime}$ (resp. $\mathrm{C}_{\sigma 2}^{\prime}$).
The following remark shows the advantage of $\mathrm{C}_{m2}^{\prime}$ over $\mathrm{C}_{m2}$.

\begin{remark}
    As stated in Remark~\ref{rmk_pseudoJ}, to fulfill $\mathrm{C}_{m2}$, the existing methods in \cite{Sun2013Weighted, LinComposite2019, Jiang2016Single} rely on the estimation of the inverse of the CDF of $\varepsilon$, which is unavailable directly from the sample set of $\sbr{X, Y}$.
    Instead of directly related to the distribution of the error, the condition $\mathrm{C}_{m2}^{\prime}$ only relies on the two expectations in~\eqref{eq_C2p}, which can be directly estimated by the sample of $\sbr{X, Y}$, and the estimation has the convergence rate of parametric estimation.
    Moreover, in the next subsection, we will see that $\mathrm{C}_{m2}^{\prime}$ can be easily expressed in a renewable form.
\end{remark}

Although the condition $\mathrm{C}_{\sigma 2}^{\prime}$ in \eqref{eq_Cs2p} relies on the unknown regression function $m\sbr{\cdot}$, it will be shown in the next subsection that the unknown $m\sbr{\cdot}$ does not bring any essential difficulties to our renewable estimation.

\subsection{Specific Renewable WCQR Estimators and Algorithms}\label{sec_specwcqr}

To construct specific $J\sbr{\cdot}$ satisfying the conditions in the last subsection,
we can obtain specific WCQR estimators for $m\sbr{x}$ and $\sigma\sbr{x}$.

Throughout this subsection, we assume that the function $W\sbr{\cdot}$ in \eqref{eq_C2p} and \eqref{eq_Cs2p} satisfies $\mathrm{Supp}\bbr{W\sbr{\cdot}} \subset I_*$, under which we can avoid estimating $r\sbr{x; J}$ with $x$ outside of $I_*$.
The simplest example of $W\sbr{\cdot}$ is that $W(x) = I\sbr{x \in I_*}$.


\subsubsection{Estimators for the Conditional Mean}\label{sec_CondiMean}

We first consider a simple case, where the model~\eqref{eq_model} is symmetric in the sense that the PDF of $\varepsilon$ is symmetric around $0$.
For symmetric models, $\sbr{\mathrm{C}_{m1}, \mathrm{C}_{m2}}$ can be easily fulfilled whenever $J(\cdot)$ is normalized and symmetric around $1/2$.
A representative example is the $\alpha$-trimmed weight function:
\begin{equation}\label{eq_Jalptim}
    J_{m, 0.5}\sbr{\tau} = 0.5 L_{\alpha} \sbr{\tau} + 0.5 U_{\alpha} \sbr{\tau},
\end{equation}
where $L_{\alpha}\sbr{\tau} = \sbr{0.5 - \alpha}^{-1} I\sbr{\alpha \leq \tau \leq 0.5}$ and $U_{\alpha}\sbr{\tau} = \sbr{0.5 - \alpha}^{-1} I\sbr{0.5 < \tau \leq 1- \alpha}$ with $\alpha \in (0, 0.5)$ selected according to Remark~\ref{rmk_ResWeight}.
With the weight function $J_{m, 0.5}\sbr{\cdot}$, the WCQR estimator $\widetilde{r}_t(x; J_{m, 0.5})$ given by \eqref{eq_widehatthetaF} is a renewable version of the naive local $\alpha$-trimmed mean (NTM) \citep[see, e.g.,][]{Bednar1984Alpha, Boente1994Local}, and in the following, we will call $\widetilde{r}_t(x; J_{m, 0.5})$ the renewable NTM or the NTM if there is no confusion.

The following remark shows that the NTM is robust to outliers but not adaptive to symmetric or asymmetric error distributions.

\begin{remark}\label{rmk_renNTM}
    The prototype of $\widetilde{r}_t(x; J_{m, 0.5})$ is the $\alpha$-trimmed mean, which has been singled out by several prominent authors as the quintessential robust estimator of location \citep[see, e.g.,][]{Bickel1975Descriptive, Stigler1977Do, Koenker2005}.
    And we can expect that $\widetilde{r}_t(x; J_{m, 0.5})$ enjoys robustness comparable to the $\alpha$-trimmed mean.
    However, since $\widetilde{r}_t(x; J_{m, 0.5})$ actually estimate the location $r\sbr{x; J_{m, 0.5}}$ rather than the conditional mean $m\sbr{x}$, the estimation consistency deeply relies on the symmetry of the error to guarantee the conditions $\sbr{\mathrm{C}_{m1}, \mathrm{C}_{m2}}$ and the equality $r\sbr{x; J_{m, 0.5}} = m\sbr{x}$.
    When the error distribution is asymmetric, the estimator $\widetilde{r}_t(x; J_{m, 0.5})$ will suffer from a non-negligible bias caused from $r\sbr{x; J_{m, 0.5}} \neq m\sbr{x}$.
\end{remark}

To address the issue mentioned in Remark~\ref{rmk_renNTM}, we should modify the $\alpha$-trimmed weight function $J_{m, 0.5}\sbr{\cdot}$, such that the conditions $\sbr{\mathrm{C}_{m1}, \mathrm{C}_{m2}}$ can be fulfilled for general error distributions.
Thus we generalize $J_{m, 0.5}\sbr{\cdot}$ into
\begin{equation}\label{eq_Jalpw}
    J_{m, w}(\tau) = \omega L_{\alpha} \sbr{\tau} + \sbr{1 - \omega} U_{\alpha} \sbr{\tau},
\end{equation}
where $w \in \bR$ is a parameter selected to satisfy the alternative condition $\mathrm{C}_{m2}^{\prime}$, i.e.,
\begin{equation}\label{eq_EW2EWY}
    w = \frac{E_{WY} - E_{WU}}{E_{WL} - E_{WU}}
\end{equation}
with
\begin{equation*}
    \begin{aligned}
        E_{WY} &= \E{W(X) Y},\\
        E_{WL} &= \int_{I_*} W\sbr{x} r \sbr{x; L_{\alpha}} f_X\sbr{x} \di x,\\
        E_{WU} &= \int_{I_*} W\sbr{x} r \sbr{x; U_{\alpha}} f_X\sbr{x} \di x.
    \end{aligned}
\end{equation*}
Since $J_{m, w}(\cdot)$ satisfies $\sbr{\mathrm{C}_{m1}, \mathrm{C}_{m2}^{\prime}}$ implying that $r\sbr{x; J_{m, w}} = m\sbr{x}$, i.e., the non-negligible bias is corrected.
In the following text, we will call $\widetilde{r}_t\sbr{x; J_{m, w}}$ the renewable bias-corrected local $\alpha$-trimmed mean (BCTM) or the BCTM if there is no confusion.

\begin{remark}\label{rmk_symmf}
    The BCTM $\widehat{r}\sbr{x; J_{m, w}}$ is an extension of the NTM $\widehat{r}\sbr{x; J_{m, 0.5}}$.
    Actually, when the model~\eqref{eq_model} is symmetric, by straightforward calculation, we have $w = 1/2$, in which case $\widehat{r}\sbr{x; J_{m, w}}$ is identical to $\widehat{r}\sbr{x; J_{m, 0.5}}$.
\end{remark}

To obtain the BCTM for streaming data, we should construct a renewable estimator for the unknown $w$.
Based on the expressions in \eqref{eq_EW2EWY} and the plug-in estimators, $w$ can be estimated by
\begin{equation}\label{eq_estw}
    \widehat{w}_t = \frac{\widehat{E}_{WY, t} - \widehat{E}_{WU, t}}{\widehat{E}_{WL, t} - \widehat{E}_{WU, t}}
\end{equation}
where
\begin{equation*}
    \begin{aligned}
        \widehat{E}_{WY, t} &= \frac{N_{t-1}}{N_t} \widehat{E}_{WY, t-1} + \frac{1}{N_t} \sum_{j=1}^{n_t} W(X_{tj}) Y_{tj}\\
        \widehat{E}_{WL, t} &= \int_{I_*} W\sbr{x} \I_l \widetilde{r}_t \sbr{x; L_{\alpha}} \I_l \widehat{f}_{X, t}\sbr{x} \di x,\\
        \widehat{E}_{WU, t} &= \int_{I_*} W\sbr{x} \I_l \widetilde{r}_t \sbr{x; U_{\alpha}} \I_l \widehat{f}_{X, t}\sbr{x} \di x
    \end{aligned}
\end{equation*}
with the initial value $\widehat{E}_{WY, 0} = 0$.
Here $\I_l \widetilde{r}_t \sbr{\cdot\;; L_{\alpha}}$ and $\I_l \widetilde{r}_t \sbr{\cdot\;; U_{\alpha}}$ are the interpolated WCQR estimator given in \eqref{eq_Iltilder}, and $\I_l f_X\sbr{\cdot}$ is the interpolated empirical PDF given by
\begin{equation}\label{eq_IlfX}
    \I_l f_X\sbr{\cdot} = \sum_{x_i \in G_*} \widehat{f}_{X, t}\sbr{x_i} L\sbr{\cdot, x_i; l, G_*}
\end{equation}
with $\widehat{f}_{X, t}\sbr{x_i}$ the renewable statistics obtained by~\eqref{eq_St} and $G_*$ the node set introduced in \eqref{eq_Gstar}.

If we omit the error caused by LPI approximation, $\widehat{w}_t$ is a consistent estimator of $w$, and we known from the Slutsky's Lemma that the estimator $\widetilde{r}_t\sbr{x; J_{m, \widehat{w}_t}}$ has the same asymptotic distribution as that of $\widetilde{r}_t\sbr{x; J_{m, w}}$.

In the following remark, we show the pros and cons of the BCTM compared with the NTM.

\begin{remark}\label{rmk_EWY}
    Compared with the NTM, the first advantage of the BCTM  is that the estimation consistency is based on the structure of the model instead of the symmetry of the error.
    Thus the BCTM is adaptive to symmetric or asymmetric error distributions.
    However this adaptiveness comes with a little costs in robustness, when estimating the parameter $\omega$.
    Actually, it is easy to check that the estimators $\widehat{E}_{W1, t}$ and $\widehat{E}_{W2, t}$ have bounded influence functions, and thus both of them are robust to outliers.
    However, the estimator $\widehat{E}_{WY, t}$ is somewhat non-robust.
    Fortunately, the unknown $E_{WY, t}$ is a scalar parameter rather than a general function, thus the estimator $\widehat{E}_{WY, t}$ can achieve $\sqrt{N_t}$-consistency, which is faster than the optimal convergence rate of nonparametric regression estimation.
    Hence the final nonparametric estimator $\widetilde{r}_t\sbr{x; J_{m, \widehat{w}_t}}$ is less susceptible to
    the weak robustness of $\widehat{E}_{WY, t}$.
    This point will be further demonstrated in our numerical studies in Section~\ref{sec_simu}.
\end{remark}

In Algorithm~\ref{alg_wcqr_mx}, we present the detail procedures to obtain the renewable BCTM for the estimation of $m\sbr{x}$ with $x$ belonging to some gird points in $I_1$.

\begin{algorithm*}
    \caption{Renewable BCTM for estimating $\bbr{m\sbr{x_i}}_{i=1}^{\bar{q}}$ with $x_i$ gird points on $\bR$}\label{alg_wcqr_mx}
    \begin{algorithmic}[1]
    \Require Set of grid points $G_* = \bbr{x_i}_{i=1}^{\bar{q}}$, kernel function $K\sbr{\cdot}$, node sets $G_{x_i}$ for $x_i \in G_*$, PLI degree $l$
    \State Set initial values $N_0 = \widehat{E}_{WY, 0} = \widehat{f}_{X, 0}\sbr{x_i} = \widehat{S}_{Y\vert x, 0}\sbr{y_{ij}} = 0$ for $y_{ij} \in G_{x_i}$ and $x_i \in G_*$
    \For{$t = 1, 2, \cdots$}
        \State Obtain the $t$-th data chunk $\D_t = \bbr{X_{tj}, Y_{tj}}_{j=1}^{n_t}$
        \State Select the $t$-th bandwidth $h_t$
        \State $N_{t} = N_{t-1} + n_t$\label{line_Nt}
        \State $\widehat{E}_{WY, t} = N_{t-1} / N_t \widehat{E}_{WY, t-1} + 1/N_t \sum_{j=1}^{n_t} W(X_{tj}) Y_{tj}$
        \For{$x_i \in G_*$}
            \State $\widehat{f}_{X, t} \sbr{x_i} = N_{t-1}/ N_t \widehat{f}_{X, t-1}\sbr{x_i} + 1/N_t \sum_{j=1}^{n_t} K_{h_t}\left(X_{tj} - x_i\right)$, where $K_h(\cdot) = 1/h K(\cdot/ h)$
            \State $\widehat{S}_{Y\vert x_i, t} \sbr{y_{ij}} = N_{t-1}/ N_t  \widehat{S}_{Y\vert x_i, t-1} \sbr{y_{ij}} + 1/N_t \sum_{j=1}^{n_t} I\sbr{Y_{tj} < y_{ij}} K_{h_t}\left(X_{tj} - x_i\right)$ for $y_{ij} \in G_{x_i}$
        \EndFor \label{line_endfor}
        \If{the estimators for $\bbr{m\sbr{x_i}}_{x_i \in G_*}$ are needed}
            \State Define the function $\I_l \widehat{f}_{X, t}\sbr{\cdot} = \sum_{x_i \in G_*} \widehat{f}_{X, t} \sbr{x_i} L\sbr{\cdot, x_i; l, G_*}$\label{line_Def}
            \State Define the function $\I_l \widehat{F}_{Y\vert x_i, t}\sbr{\cdot} = \sum_{y_{ij} \in G_{x_i}} {\widehat{S}_{Y\vert x_i, t} \sbr{y_{ij}} /\widehat{f}_{X, t} \sbr{x_i}} L\sbr{\cdot, y_{ij}; l, G_{x_i}}$ for $x_i \in G_*$
            \For{$J \in \bbr{L_{\alpha}, U_{\alpha}}$}
                \State $\widetilde{r}_t(x_i; J) = \int_{\bR} y J(\I_l \widehat{F}_{Y\vert x_i, t}\sbr{y}) \di \I_l \widehat{F}_{Y\vert x_i, t}\sbr{y}$ for $x_i \in G_*$
                \State Define the function $\I_l \widetilde{r}_t \sbr{\cdot\;; J} = \sum_{x_i \in G_*} \widetilde{r}_t(x_i; J) L\sbr{\cdot, x_i; l, G_*}$
            \EndFor
            \State $\widehat{E}_{WL, t} = \int_{I_*} W\sbr{x} \I_l \widetilde{r}_t \sbr{x; L_{\alpha}} \I_l \widehat{f}_{X, t}\sbr{x} \di x$
            \State $\widehat{E}_{WU, t} = \int_{I_*} W\sbr{x} \I_l \widetilde{r}_t \sbr{x; U_{\alpha}} \I_l \widehat{f}_{X, t}\sbr{x} \di x$
            \State $\widehat{w}_t = \sbr{\widehat{E}_{WY, t} - \widehat{E}_{WU, t}}/\sbr{\widehat{E}_{WL, t} - \widehat{E}_{WU, t}}$
            \State $\widetilde{r}_t(x_i; J_{m, \widehat{w}_t}) = \widehat{w}_t \widetilde{r}_t(x_i; L_{\alpha}) + \sbr{1-\widehat{w}_t}\widetilde{r}_t(x_i; U_{\alpha})$ for $x_i \in G_*$\label{line_Out}
            \State Output $\bbr{\widetilde{r}_t(x_i; J_{m, \widehat{w}_t})}_{x_i \in G_*}$ as the estimators for $\bbr{m\sbr{x_i}}_{x_i \in G_*}$
        \EndIf
    \EndFor
    \end{algorithmic}
\end{algorithm*}

\subsubsection{Estimators for the Conditional Variance}

To estimate $\sigma\sbr{x}$, the first condition $\mathrm{C}_{\sigma 1}$ can be fulfilled by taking $J\sbr{\cdot}$ antisymmetry around $1/2$.
Then parallel to the NTM, we introduce the antisymmetry $\alpha$-trimmed weight function:
\begin{equation}\label{eq_Jsgmthata}
    J_{\sigma, 1}(\tau) = \sbr{- L_{\alpha} \sbr{\tau} + U_{\alpha} \sbr{\tau}}.
\end{equation}
From \eqref{dec_rxJ} and \eqref{eq_Jsgmthata}, we can see that
\begin{equation*}
    r\sbr{x; J_{\sigma, 1}} = c_0 \sigma(x) \sptext{for} x \in I_*
\end{equation*}
with $c_0 = \int_{[0, 1]} J_{\sigma, 1}(\tau) F_{\varepsilon}^{-1}\sbr{\tau} \di \tau$ a constant independent with $x$.
Thus the renewable WCQR estimator $\tilde{r}_t\sbr{x; J_{\sigma, 1}}$ is qualified to estimate $\sigma\sbr{x}$ consistently up to scale; that is,
\begin{equation*}
    \tilde{r}_t \sbr{x; J_{\sigma, 1}} = c_0 \sigma(x) + O\sbr{\abs{\Delta\sbr{G_x}}^{l+1}} + o_p\sbr{1}.
\end{equation*}
In the following text, we call $\tilde{r}_t \sbr{x; J_{\sigma, 1}}$ the naive $\alpha$-trimmed standard derivation (NTSD), since it is modified from the NTM and is used to estimate the conditional standard derivation.

If we aim higher and want to estimate $\sigma\sbr{x}$ consistently, we should rescale the weight function such that $c_0 = 1$, or equivalently, $\mathrm{C}_{\sigma 2}$ holds.
To this end, we extend $J_{\sigma, 1}(\cdot)$ into
\begin{equation}\label{eq_Jsigtheta}
    J_{\sigma, \theta}(\tau) = \theta J_{\sigma, 1}(\tau)
\end{equation}
where $\theta$ is a scale parameter selected to satisfying the alternative $\mathrm{C}_{\sigma 2}^{\prime}$, i.e.,
\begin{equation}\label{eq_theta2}
    \theta^2 = \frac{\E{W(X) \sbr{Y^2 - m^2 \sbr{X}}}}{\E{W(X) r^2\sbr{X; J_{\sigma, 1}}}}.
\end{equation}
The main issue is to obtain a renewable estimation of the unknown $\theta$.
Similar to the idea introduced in Section~\ref{sec_CondiMean}, we can estimate $\theta$ by
\begin{equation}\label{eq_hatthetat}
    \widehat{\theta}_t = \sqrt{\frac{\widehat{E}_{WY^2, t} - \widehat{E}_{Wm^2, t}}{\widehat{E}_{Wr^2, t}}}
\end{equation}
where
\begin{equation*}
\begin{aligned}
    \widehat{E}_{WY^2, t} &= \frac{N_{t-1}}{N_t} \widehat{E}_{WY^2, t-1} + \frac{1}{N_t} \sum_{j=1}^{n_t} W\sbr{X_{tj}} Y_{tj}^2,\\
    \widehat{E}_{Wm^2, t} &= \int_{I_*} W\sbr{x} \widetilde{m}_t^2 (x) \I_l \widehat{f}_{X, t}\sbr{x} \di x,\\
    \widehat{E}_{Wr^2, t} &= \int_{I_*} W\sbr{x} \sbr{\I_l \widetilde{r}_t}^2(x; J_{\sigma, 1}) \I_l \widehat{f}_{X, t}\sbr{x} \di x
\end{aligned}
\end{equation*}
with the initial value $\widehat{E}_{WY^2, 0} = 0$.
Here $\I_l \widetilde{r}_t \sbr{\cdot; J_{\sigma, 1}}$ and $\I_l \widehat{f}_{X, t}\sbr{\cdot}$ are renewable estimator obtained by \eqref{eq_Iltilder} and \eqref{eq_IlfX}, respectively, and $\widetilde{m}_t (\cdot)$ is an renewable estimator of $m\sbr{\cdot}$, e.g.,
\begin{equation*}
    \widetilde{m}_t (\cdot) = \I_l \widetilde{r}_t (\cdot\;; J_{m, w}).
\end{equation*}
with $w = 0.5$ for symmetric models and $w = \widehat{w}_t$ given in \eqref{eq_estw} for asymmetric models.

Based on \eqref{eq_widehatthetaF} with $J_{\sigma, \widehat{\theta}_t}\sbr{\cdot}$ in place of $J\sbr{\cdot}$, we can obtain the renewable estimator $\widetilde{r}_t (x; J_{\sigma, \widehat{\theta}_t})$ for the conditional standard deviation $\sigma\sbr{x}$.
In the following text, we will call $\widetilde{r}_t (x; J_{\sigma, \widehat{\theta}_t})$ the rescaled $\alpha$-trimmed standard derivation (RTSD).

In the following remark, we make a comparison between the NTSD and the RTSD.
\begin{remark}
    The RTSD $\widetilde{r}_t (x; J_{\sigma, \widehat{\theta}_t})$ contains more information about $\sigma\sbr{x}$, since it can identify the constant $c_0$, which is unrevealed by the NTSD $\tilde{r}_t \sbr{x; J_{\sigma, 1}}$.
    In another aspect, the NTSD enjoys desirable robustness comparable to the classic $\alpha$-trimmed mean.
    While the RTSD depends on the plug-in estimator $\widehat{E}_{WY^2, t}$, and similar to the discussions in Remark~\ref{rmk_EWY},
    $\widehat{E}_{WY^2, t}$ is somewhat non-robust but enjoys the convergence rate of parametric estimation.
\end{remark}

In the supplementary material, we present a complete algorithm to obtain renewable WCQR estimators for $m\sbr{x}$ and $\sigma\sbr{x}$ from asymmetric models.

In the following remark, we show the desirable computation efficiency of our algorithms in dealing with rapid data steams.
\begin{remark}
    Algorithm~\ref{alg_wcqr_mx} mainly consists of two parts: the updating part (Lines~\ref{line_Nt} - \ref{line_endfor}) where the cumulative statistics are updated, and the estimation part (Lines~\ref{line_Def} - \ref{line_Out}) where the WCQR estimator is computed by calculating integrals.
    Notice that the updating part is relatively simple without solving any nonlinear equations.
    And at each updating step, the computations among each $x_* \in G_*$ can be implemented parallelly.
    Benefit from this, the updating part can be implemented fast enough to catch up with the rapid data steam.
    The estimation part seems to be computationally intensive.
    Fortunately, this part is implemented only when one needs the current value of the WCQR estimator.
    Thus it would not cause trouble in computation speed even when the data stream is rapid.
    The above desirable feature is also enjoyed by Algorithm A.1 in the supplementary material.
\end{remark}

\section{Theoretical Analyses}\label{sec_theo}

In this section, we will deduce the asymptotic distribution of the renewable WCQR estimator, based on which we will propose renewable bandwidth selectors and obtain the optimal weight functions by minimizing the asymptotic variance.

\subsection{Asymptotic Distributions}

For theoretical analyses of the estimator $\widetilde{r}_t\sbr{x; J}$, we introduce the following standard assumptions.

\begin{assumption}\label{assum_fmsigma}
    The functions $f_X(\cdot)$, $m(\cdot)$ and $\sigma(\cdot)$ have continuous derivatives up to order $4$ on $I_*$.
    On a open set containing $\mathrm{Supp}\bbr{J\sbr{\cdot}}$, the function $F_{\varepsilon}(\cdot)$ has continuous derivatives up to order $\max\bbr{l+1, 4}$, where $l$ is the degree of PLI given in \eqref{eq_intpFYx}.
    The function $f_X(\cdot)$ admits a positive lower bounded on~$I_*$.
\end{assumption}
\begin{assumption}\label{assum_K}
    The kernel function $K(u)$ is symmetric and compactly supported, and satisfies that $\int_{\bR} K(u) \di u=1$, $k_{4, 1} = \int_{\bR} u^4 K(u) \di u < \infty$ and $k_{0, 2} = \int_{\bR} K^2(u) \di u < \infty$.
\end{assumption}
\begin{assumption}\label{assum_htbt}
    The bandwidth $h_t$ and smoothing parameter $b_t$ satisfies that as $t \to \infty$,
    \begin{equation}\label{eq_ht0}
        h_t = o(1) \sptext{and} \sum_{s=1}^t \frac{n_s}{h_s} \sbr{\sum_{s=1}^t n_s h_s^4}^{-2} = o\sbr{1},
    \end{equation}
    \begin{equation}\label{eq_htbt}
        \frac{1}{N_t} \sum_{s=1}^t \frac{n_s}{N_t h_s} = o(1).
    \end{equation}
\end{assumption}
\begin{assumption}\label{assum_J}
    The L-score function $J(\cdot)$ is piecewise continuously differentiable.
    There exist constants $0 < \utau < \btau < 1$ such that $\mathrm{Supp}\bbr{J\sbr{\cdot}} \subset [\utau, \btau]$.
    The node set $G_x$ satisfies $[\utau, \btau] \subset (\min F_{Y\vert x}\sbr{G_x}, \max F_{Y\vert x}\sbr{G_x})$.
\end{assumption}

In Assumptions~\ref{assum_fmsigma} and \ref{assum_K}, all the conditions on $f_X(\cdot)$, $m(\cdot)$, $\sigma(\cdot)$ and $K\sbr{\cdot}$ are standard for nonparametric regressions, and the condition on $F_{\varepsilon}(\cdot)$ is required by the LPI approximation to guarantee its accuracy.
Among the conditions in Assumption~\ref{assum_htbt}, the first one in \eqref{eq_ht0} is required for a vanishing estimation bias; the second one in \eqref{eq_ht0} is used to simplify the bias terms in the asymptotic results; the last condition \eqref{eq_htbt} is necessary for a bounded estimation variance, and it is fulfilled whenever $N_t \min_{1 \leq s\leq t} h_s \to \infty$.
Assumption~\ref{assum_J} is used to bound the remainder term in the asymptotic expansion of $\widetilde{r}_t\sbr{x; J}$; the restrictions on $\mathrm{Supp}\bbr{J\sbr{\cdot}}$ and $G_x$ are also required by the LPI approximation and in practical applications, it can be fulfilled empirically; see Remark~\ref{rmk_ResWeight}.


The following theorem gives the asymptotic properties of the interpolated empirical CDF $\I_l \widehat{F}_{Y\vert x, t}\sbr{\cdot}$.

\begin{theorem}\label{thm_asy_Fyx}
    Under Assumptions~\ref{assum_fmsigma} - \ref{assum_htbt}, it holds for $y \in \bR$ that
    \begin{equation*}
        \begin{aligned}
            &\bbr{\frac{1}{N_t} \sum_{s=1}^t \frac{n_s}{N_t h_s}}^{-1/2} \bbr{\I_l \widehat{F}_{Y\vert x, t}\sbr{y} - F_{Y\vert x}\sbr{y}\right. \\
            &\left. - R_{\I_l, F, x}\sbr{y} - \I_l B_{F, x}\sbr{y} \sum_{s=1}^t \frac{n_s h_s^2}{N_t}} \\
            &\overset{\mathrm{d}}{\to} N\sbr{0, \I_l^2 C_{F, x} \sbr{y, y}},
        \end{aligned}
    \end{equation*}
    where
    \begin{align*}
        &R_{\I_l, F, x}\sbr{y} = -\frac{F_{Y\vert x}^{(l+1)}\sbr{y}(c)}{\sbr{l+1}!}\prod_{y_i \in \mathcal{N}_{l+1}\sbr{y, G_x}} \sbr{y - y_i}, \\
        &\I_l B_{F, x}\sbr{y} = \sum_{y_i \in G_x} B_{F, x}\sbr{y_i} L\sbr{y, y_i; l, G_x},
    \end{align*}
    \begin{equation*}
        \begin{aligned}
            \I_l^2 C_{F, x} \sbr{z_1, z_2} =& \sum_{y_i, y_j \in G_x} L\sbr{z_1, y_i; l, G_x} \times \\
            & L\sbr{z_2, y_j; l, G_x} C_{F, x}\sbr{y_i, y_j}
        \end{aligned}
    \end{equation*}
    with
    \begin{align}
        B_{F, x}\sbr{y_i} =\;& \frac{k_{2, 1}}{2} \bbr{\partial_x^2 F_{Y \vert x}\sbr{y_i}\right. \notag\\
        &\left. + 2 \partial_x F_{Y \vert x}\sbr{y_i} \frac{f_X^{\prime}\sbr{x}}{f_X\sbr{x}}}, \label{eq_BFx} \\
        C_{F, x}\sbr{y_i, y_j} =\;& \frac{k_{0, 2}}{f_X\sbr{x}}\bbr{F_{Y \vert x}\sbr{y_i \land y_j}\right.\notag \\
        &\left. - F_{Y \vert x}\sbr{y_i} F_{Y \vert x}\sbr{y_j}} \label{eq_CFx}
    \end{align}
    and $c \in \rbr{\min \overline{\mathcal{N}}_{l+1}\sbr{y}, \max \overline{\mathcal{N}}_{l+1}\sbr{y}}$ with $\overline{\mathcal{N}}_{l+1}\sbr{y}$ given in Lemma~\ref{lemm_LPI}.
\end{theorem}

Based on Theorem~\ref{thm_asy_Fyx}, we can conclude that
\begin{equation}\label{eq_normIlFYxtO}
    \begin{aligned}
        \norm{\I_l \widehat{F}_{Y\vert x, t} - F_{Y\vert x}}_{2, I\sbr{G_x}} = O(\abs{\Delta \sbr{G_x}}^{l+1}) \\
        + O\sbr{\sum_{s=1}^t \frac{n_s h_s^2}{N_t}} +  O_p\sbr{\sqrt{\frac{1}{N_t} \sum_{s=1}^t \frac{n_s}{N_t h_s}}}.
    \end{aligned}
\end{equation}
From \eqref{eq_normIlFYxtO} we can see that the error of our interpolated empirical CDF consists of two parts: the first part depending  on $\Delta \sbr{G_x}$ is a numerical error caused from the LPI approximation, and the second part depending on the bandwidths consists of statistical errors corresponding to the bias and variance of kernel estimations.
Because the statistical errors have a convergence rate the same with that of the oracle empirical CDF obtained on the imaginary full data set $\cup_{s \leq t} \D_t$,
and the numerical error is usually negligible compared with the statistical ones (see the following Remark~\ref{rmk_numsta_error}),
it can be expected that our estimator enjoys a performance almost as well as the oracle estimator.

\begin{remark}\label{rmk_numsta_error}
    In practical applications, the numerical error in \eqref{eq_normIlFYxtO} is usually much smaller than the remaining statistical errors.
    Actually, when the nodes in $G_x$ are uniformly spaced, the numerical error is of order $O(\sbr{\# G_x}^{-(l+1)})$, which can be reduced significantly by increasing the number $\#G_x$ of nodes and the degrees $l$ of LPI, whenever the computation resources permit.
    Unlike the numerical one, the statistical errors deeply rely on the number $N_t$ of samples, and the convergence rate is relatively slow (no more than $O(N_t^{-2/5})$).
\end{remark}

Based on Theorem~\ref{thm_asy_Fyx}, we have the following main result on the asymptotic property of the estimator $\widetilde{r}_t(x; J)$.

\begin{theorem}\label{thm_asytheta}
    Under Assumptions~\ref{assum_fmsigma} - \ref{assum_J}, it holds that
    \begin{equation}\label{eq_asy_tildemx}
        \begin{aligned}
            &\bbr{\frac{1}{N_t} \sum_{s=1}^t \frac{n_s}{N_t h_s}}^{-1/2} \bbr{\widetilde{r}_t(x; J) - r(x; J) \right.\\
            &\left.- B_{\I_l, m, x} \sum_{s=1}^t \frac{n_s h_s^2}{N_t} - R_{\I_l, m, x}} \overset{\mathrm{d}}{\to} N\sbr{0, \Sigma_{\I_l, m, x}}.
        \end{aligned}
    \end{equation}
    where
    \begin{equation*}
        \begin{aligned}
            B_{\I_l, m, x} =\;& -\int_{\bR} J\sbr{F_{Y\vert x}\sbr{y}} \I_l B_{F, x}\sbr{y} \di y,\\
            R_{\I_l, m, x} =\;& -\int_{\bR} J\sbr{F_{Y\vert x}\sbr{y}} R_{\I_l, F, x}\sbr{y} \di y \\
            &+ O\sbr{\abs{\Delta \sbr{G_x}}^{2l + 2}}, \\
            \Sigma_{\I_l, m, x} =\;& \int_{\bR}\int_{\bR} C_{\I_l, J, x}\sbr{z_1, z_2} \di z_1 \di z_2\\
        \end{aligned}
    \end{equation*}
    with
    \begin{equation*}
        \begin{aligned}
            C_{\I_l, J, x}\sbr{z_1, z_2} =\;& J\sbr{F_{Y\vert x}\sbr{z_1}}J\sbr{F_{Y\vert x}\sbr{z_2}} \\
            &\times \I_l^2 C_{F, x} \sbr{z_1, z_2}
        \end{aligned}
    \end{equation*}
    and $\I_l B_{F, x}\sbr{y}$, $R_{\I_l, F, x}\sbr{y}$ and $\I_l^2 C_{F, x} \sbr{z_1, z_2}$ given in Theorem~\ref{thm_asy_Fyx}.
\end{theorem}

Theorem~\ref{thm_asytheta} shows that the LPI operator $\I_l$ has influence on the asymptotic bias and variance of $\widetilde{r}_t(x; J)$, while its influence is no more than the order of numerical errors.
The following corollary states the details.

\begin{corollary}\label{coro_dom_terms}
    Under the conditions of Theorem~\ref{thm_asytheta}, it holds that
    \begin{equation*}
        \begin{aligned}
            B_{\I_l, m, x} =\;& B_{m, x} + O\sbr{\abs{\Delta \sbr{G_x}}^{l+1}},\\
            R_{\I_l, m, x} =\;& O\sbr{\abs{\Delta \sbr{G_x}}^{l+1}}\\
            \Sigma_{\I_l, m, x} =\;& \Sigma_{m, x} + O\sbr{\abs{\Delta \sbr{G_x}}^{l+1}},
        \end{aligned}
    \end{equation*}
    where the dominant terms are
    \begin{equation*}
        \begin{aligned}
            B_{m, x} =\;& -\int_{\bR} J\sbr{F_{Y\vert x}\sbr{y}} B_{F, x}\sbr{y} \di y\\
            \Sigma_{m, x} =\;& \frac{k_{0, 2} \sigma^2 (x)}{f_X(x)} \int_{[\utau, \btau]} \int_{[\utau, \btau]} S_J\sbr{\tau_1, \tau_2} \di \tau_1 \di \tau_2
        \end{aligned}
    \end{equation*}
    with
    \begin{equation}
        S_J\sbr{\tau_1, \tau_2} = \frac{\sbr{\tau_1 \land \tau_2 - \tau_1 \tau_2} J\sbr{\tau_1} J\sbr{\tau_2}}{f_{\varepsilon}\sbr{F_{\varepsilon}^{-1}\sbr{\tau_1}} f_{\varepsilon}\sbr{F_{\varepsilon}^{-1}\sbr{\tau_2}}}.
    \end{equation}
\end{corollary}

Combining Theorem~\ref{thm_asytheta} and Corollary~\ref{coro_dom_terms}, we can find that the error between  $\widetilde{r}_t(x; J)$ and $\widetilde{r}_t(x; J)$ has a convergence rate the same with the error given in \eqref{eq_normIlFYxtO}, i.e.,
\begin{equation}\label{eq_converge_tildert}
    \begin{aligned}
        &\widetilde{r}_t(x; J) - r(x; J) = O(\abs{\Delta \sbr{G_x}}^{l+1}) \\
        &\quad + O\sbr{\sum_{s=1}^t \frac{n_s h_s^2}{N_t}} +  O_p\sbr{\sqrt{\frac{1}{N_t} \sum_{s=1}^t \frac{n_s}{N_t h_s}}},
    \end{aligned}
\end{equation}
where the first term is the numerical error caused from LPI, and the remaining two terms are statistical errors caused from kernel estimations.
By the discussion in Remark~\ref{rmk_numsta_error}, the numerical error is usually negligible compared with the statistical ones.
Moreover, the convergence rates of the statistical errors are   the same with that of the oracle estimator $\widehat{r}(x; J)$ obtained on the imaginary full data set $\cup_{s \leq t} \D_t$.
Theoretically, our renewable WCQR estimator behaves almost as well as the oracle estimator.

Based on the relation \eqref{dec_rxJ} and the results in Lemma~\ref{lemm_intJF0} and Theorem~\ref{thm_asytheta}, we can immediately obtain the following corollary, which gives asymptotic properties of the estimators introduced in Section~\ref{sec_specwcqr}.

\begin{corollary}
    Given $J\sbr{\cdot}$ satisfying $(\mathrm{C}_{m1}, \mathrm{C}_{m2})$ or $(\mathrm{C}_{m1}, \mathrm{C}_{m2}^{\prime})$ (resp. $(\mathrm{C}_{\sigma 1}, \mathrm{C}_{\sigma 2})$ or $(\mathrm{C}_{\sigma 1}, \mathrm{C}_{\sigma 2}^{\prime})$), the result \eqref{eq_asy_tildemx} in Theorem~\ref{thm_asytheta} holds with $r\sbr{x; J}$ replaced with $m\sbr{x}$ (resp. $\sigma\sbr{x}$).
\end{corollary}

\subsection{The Selection of Bandwidths}\label{sec_selectbw}

For implementing bandwidth selection, we need to calculate the asymptotic mean square error (AMSE) and the asymptotic mean integrated squared error (AMISE) between $\widetilde{r}_t(x; J)$ and $r(x; J)$.
It follows from Theorem~\ref{thm_asytheta} and Corollary~\ref{coro_dom_terms} that the asymptotic bias and variance of $\widetilde{r}_t(x; J)$ can be given by
\begin{align*}
    &\mathrm{Bias}\bbr{\widetilde{r}_t(x; J)} = B_{m, x} \sum_{s=1}^t \frac{n_s}{N_t} h_s^2 \\
    &\quad\quad + o\sbr{\sum_{s=1}^t \frac{n_s}{N_t} h_s^2} + O(\abs{\Delta \sbr{G_x}}^{l+1}),\\
    &\mathrm{Var}\bbr{\widetilde{r}_t(x; J)} = \frac{1}{N_t}\sum_{s=1}^t \frac{n_s}{N_t h_s} \Sigma_{m, x}\\
    &\quad\quad + \frac{1}{N_t}\sum_{s=1}^t \frac{n_s}{N_t h_s} \sbr{o\sbr{1} + O(\abs{\Delta \sbr{G_x}}^{l+1})}.
\end{align*}
By Remark~\ref{rmk_numsta_error}, it is reasonably to assume  the numerical error is negligible compared with the statistical ones, thus we can omit the terms $O(\abs{\Delta \sbr{G_x}}^{l+1})$ and other high order terms in the asymptotic bias and variance.
Then the AMSE and AMISE can be respectively defined as
\begin{equation*}
    \begin{aligned}
        \mathrm{AMSE} \bbr{\widetilde{r}_t(x; J)} &= \mathcal{E}_t \sbr{x; H_t}, \\
        \mathrm{AMISE} \bbr{\widetilde{r}_t(x; J)} &= \int_{I_*} \mathcal{E}_t \sbr{x; H_t} \widetilde{W}(x) \di x,
    \end{aligned}
\end{equation*}
where $\widetilde{W}(\cdot)$ is a weight function defined on $I_*$;
$H_t = \sbr{h_1, \cdots, h_t}^{\top}$ is a vector consisting of the used bandwidth components $h_s$ up to the $t$-th data chunk and $\mathcal{E}_t \sbr{x; \cdot}$ is an error function defined by
\begin{equation}\label{eq_Etxh}
    \begin{aligned}
        \mathcal{E}_t \sbr{x; H_t} =\;& \sbr{B_{m, x} \sum_{s=1}^t \frac{n_s h_s^2}{N_t}}^2 \\
        &+ \frac{1}{N_t}\sum_{s=1}^t \frac{n_s}{N_t h_s} \Sigma_{m, x}
    \end{aligned}
\end{equation}
with the asymptotic bias $B_{m, x}$ and the asymptotic variance $\Sigma_{m, x}$ given in Corollary~\ref{coro_dom_terms}.

\subsubsection{Theoretically optimal bandwidths}

We first consider an ideal situation where the streaming data is finite with the terminal time $T$ and the cumulative number $N_T$ known throughout the updating procedures $t = 1, \cdots, T$.
In this case, the theoretical optimal variable bandwidths $\tilde{H}_T^*\sbr{x} = (\tilde{h}_1^*\sbr{x}, \cdots, \tilde{h}_T^*\sbr{x})^{\top}$ and the optimal constant bandwidth $\tilde{H}_T^* = (\tilde{h}_1^*, \cdots, \tilde{h}_T^*)^{\top}$ can be obtained conventionally by minimizing the terminal AMSE and AMISE, respectively, i.e.,
\begin{equation*}
    \begin{aligned}
        &\tilde{H}_T^* \sbr{x} = \argmin_{H_T \in [0, \infty)^{T}} \mathcal{E}_T \sbr{x; H_T}, \\
        &\tilde{H}_T^* = \argmin_{H_T \in [0, \infty)^{T}} \int_{I_*} \mathcal{E}_T \sbr{x; H_T} \widetilde{W}(x) \di x.
    \end{aligned}
\end{equation*}
A straightforward calculation leads to
\begin{equation}\label{eq_optht}
    \tilde{h}_t^*\sbr{x} = \sbr{C_h\sbr{x}}^{1/5} N_T^{-1/5}, \; \tilde{h}_t^* = \sbr{C_h}^{1/5} N_T^{-1/5}
\end{equation}
for $t = 1, \cdots, T$, where
\begin{equation*}
\begin{aligned}
    &C_h\sbr{x} = \frac{\Sigma_{m, x}}{4 \sbr{B_{m, x}}^2}, \\
    &C_h = \frac{\int_{I_*}\Sigma_{m, x} \widetilde{W}(x) \di x}{4 \int_{I_*} \sbr{B_{m, x}}^2 \widetilde{W}(x) \di x}.
\end{aligned}
\end{equation*}

The following remark shows the standard statistical convergence rate of $\widetilde{r}_T(x; J)$ under the optimal bandwidths.
\begin{remark}
    Recalling the error expressions in \eqref{eq_normIlFYxtO} and \eqref{eq_converge_tildert}, we conclude that
    when the bandwidths are the optimal ones given in \eqref{eq_optht},
    the statistical errors of $\I_l \widehat{F}_{Y\vert x, T}\sbr{\cdot}$ and $\widetilde{r}_T\sbr{x; J}$ both enjoy the optimal convergence rate of $O_p\sbr{N_T^{-2/5}}$.
    We should remark that such a standard convergence rate is obtained on the streaming data sets without any restrictions on the chunk size or chunk number.
\end{remark}

\subsubsection{Practical sub-optimal bandwidths}

It is usually closer to the real condition that the streaming data are endless or the terminal cumulative number $N_T$ is unpredictable.
In such a case, the theoretical optimal bandwidths in \eqref{eq_optht} are impractical, and we have to find sub-optimal bandwidths that do not rely on the information of future streaming data sets.
To this end, we introduce the following lemma, which reveals the structure of the error function given in \eqref{eq_Etxh}.

\begin{lemma}\label{lemm_decEt}
    The function $\mathcal{E}_t$ defined in \eqref{eq_Etxh} has the decomposition:
    \begin{equation*}
        \begin{aligned}
            \mathcal{E}_t \sbr{x; H_t} =\;& \sbr{\frac{N_{t-1}}{N_t}}^2 \mathcal{E}_{t-1} \sbr{x; H_{t-1}} \\
            &+ \sbr{\frac{n_t}{N_t}}^2 E_t \sbr{x, h_t; H_{t-1}}
        \end{aligned}
    \end{equation*}
    with $E_t \sbr{x, h_t; H_{t-1}}$ defined by
    \begin{equation*}
        \begin{aligned}
            E_t \sbr{x, h_t; H_{t-1}}=\;& h_t^4  \sbr{B_{m, x}}^2 + \frac{1}{n_t h_t} \Sigma_{m, x} \\
            &+ 2 h_t^2 \sum_{s=1}^{t-1} \frac{n_s h_s^2}{n_t} \sbr{B_{m, x}}^2.
        \end{aligned}
    \end{equation*}
\end{lemma}

From Lemma~\ref{lemm_decEt}, we can see that at $t$-th updating procedure, the AMSE $\mathcal{E}_t$ can be expressed as a weighted sum of two parts: the first one $\mathcal{E}_{t-1}$ is the error of the last updating step, which has a fixed value at the current updating procedure; the second part $E_t \sbr{x, h_t; H_{t-1}}$ relies on $h_t$ and $H_{t-1}$, where $H_{t-1}$ has been given in the previous updating procedure and only $h_t$ is need to be determined.
Thus the optimal value of $h_t$ should minimize $E_t \sbr{x, h_t; H_{t-1}}$ after the value of $H_{t-1}$ is given.
We thus define the sub-optimal variable and constant bandwidths respectively by
\begin{equation*}
    \begin{aligned}
        &h_t^*\sbr{x} = \argmin_{h_t \in [0, \infty)} E_t \sbr{x, h_t; H_{t-1}^*\sbr{x}}, \\
        &h_t^* = \argmin_{h_t \in [0, \infty)} \int_{I_*} E_t \sbr{x, h_t; H_{t-1}^*} \widetilde{W}(x) \di x,
    \end{aligned}
\end{equation*}
where $H_{t-1}^*\sbr{x} = \sbr{h_1^*\sbr{x}, \cdots, h_{t-1}^*\sbr{x}}^{\top}$ and $H_{t-1}^* = \sbr{h_1^*, \cdots, h_{t-1}^*}^{\top}$ are the sub-optimal bandwidths selected for the first $t-1$ data chunks.
By straightforward calculation, the sub-optimal bandwidths can be obtained by solving the following equations:
\begin{equation}\label{eq_htx}
    \begin{aligned}
        h_t^*\sbr{x} =\;& \bbr{C_h\sbr{x}}^{1/3} \sbr{\sum_{s=1}^{t-1} n_s \sbr{h_s^*\sbr{x}}^2 \right.\\
        &\left. + n_t \sbr{h_t^*\sbr{x}}^2}^{-1/3},
    \end{aligned}
\end{equation}
\begin{equation}\label{eq_ht}
    h_t^* = C_h^{1/3} \sbr{\sum_{s=1}^{t-1} n_s \sbr{h_s^*}^2 + n_t \sbr{h_t^*}^2}^{-1/3}.
\end{equation}

The sub-optimal bandwidths in \eqref{eq_htx} and \eqref{eq_ht} are given by non-linear equations, which are not convenient for practical applications, especially in the scenario of the fast online-updating.
To address this issue, we use $h_{s-1}^*\sbr{x}$ and $h_{s-1}^*$ to approximate the unknown $h_s^*\sbr{x}$ and $h_s^*$ on the right side of \eqref{eq_htx} and \eqref{eq_ht}, respectively, which lead to the renewable bandwidths that
\begin{equation}\label{eq_subhtx}
\begin{aligned}
    &\widehat{h}_t\sbr{x} =  \bbr{C_h\sbr{x}}^{1/3} \bbr{S_{h, t}\sbr{x}}^{-1/3}, \\
    &\widehat{h}_t =  C_h^{1/3} S_{h, t}^{-1/3}
\end{aligned}
\end{equation}
for $t = 2, \cdots, T$, where
\begin{equation*}
\begin{aligned}
    &S_{h, t}\sbr{x} = S_{h, t-1}\sbr{x} + n_{t} \sbr{\widehat{h}_{t-1}^*\sbr{x}}^2, \\
    &S_{h, t-1} = S_{h, t-1} + n_{t} \sbr{\widehat{h}_{t-1}^*}^2
\end{aligned}
\end{equation*}
with the initial values given by $S_{h, 1}\sbr{x} = S_{h, 1} = 0$, $\tilde{h}_1^*\sbr{x} = \sbr{C_h\sbr{x}}^{1/5} n_1^{-1/5}$ and $\tilde{h}_t^* = \sbr{C_h}^{1/5} n_1^{-1/5}$.

The optimal and sub-optimal bandwidths in \eqref{eq_optht} and \eqref{eq_subhtx} all depend on unknown parameters.
In practical applications, the unknown parameters can be estimated by cross validations on a validation data set.
More details will be discussed in Section~\ref{sec_simu}.

\subsection{The Optimal Weight Functions}

Although the renewable WCQR estimators introduced in Section~\ref{sec_specwcqr} can estimate $m\sbr{x}$ and $\sigma\sbr{x}$ at a convergence rate comparable to the oracle estimators, their weight functions are generally not optimal in terms of estimation variance.
Based on the asymptotic distribution in Theorem~\ref{thm_asytheta}, we can go a step further to find the optimal weight function in the sense the associated renewable estimator  estimate $m\sbr{x}$ or $\sigma\sbr{x}$ with a minimized asymptotic variance.

To this end, we introduce the set $\mathbb{J}_m\sbr{\utau, \btau}$ (resp. $\mathbb{J}_{\sigma}\sbr{\utau, \btau}$) consisting of all the weight functions $J(\cdot): [0, 1] \to \bR$ satisfying the constrains $\sbr{\mathrm{C}_{m1}, \mathrm{C}_{m2}}$ (resp. $\sbr{\mathrm{C}_{\sigma 1}, \mathrm{C}_{\sigma 2}}$) and being square integrable with a support in $[\utau, \btau] \subset (0, 1)$.
Here $\utau$ and $\btau$ are pre-given parameters introduced in Assumption~\ref{assum_J}.
By Theorem~\ref{thm_asytheta} and Corollary~\ref{coro_dom_terms}, the variance of $\widehat{r}(x; J)$ can be expressed as
\begin{equation*}
    \begin{aligned}
        &\sbr{\frac{1}{N_t} \sum_{s=1}^t \frac{n_s}{N_t h_s}}^{-1}\text{Var}\bbr{\tilde{m}(x; J)} \\
        =\;& \frac{k_{0, 2} \sigma^2 (x)}{f_X(x)} V\sbr{J} + O\sbr{\abs{\Delta \sbr{G_x}}^{l+1}} + o\sbr{1}
    \end{aligned}
\end{equation*}
with $V\sbr{\cdot}$ a quadratic functional given by
\begin{equation}\label{eq_defVJ}
    V\sbr{J} = \int_{[\utau, \btau]} \int_{[\utau, \btau]} S_J\sbr{\tau_1, \tau_2} \di \tau_1 \di \tau_2.
\end{equation}
For $z\sbr{\cdot} = m\sbr{\cdot}$ or $\sigma\sbr{\cdot}$, the optimal weight function $J_z^*(\cdot)$ for estimating $z\sbr{x}$ is given by the following functional minimization problem:
\begin{equation}\label{eq_optJ}
    J_z^*\sbr{\cdot} = \argmin_{J \in \mathbb{J}_z\sbr{\utau, \btau}} V\sbr{J} \sptext{for} z = m, \sigma.
\end{equation}
By tools of variational analysis, we can obtain the closed-form expression of $J_z^*\sbr{\cdot}$ as in the following theorem and corollary.

\begin{theorem}\label{thm_optJ}
    Given $f_{\varepsilon}\sbr{\cdot}$ twice differentiable, the optimal weight function in \eqref{eq_optJ} can be expressed by
    \begin{equation}\label{eq_Jopt}
        J_z^*\sbr{\tau} = I\sbr{\utau \leq \tau \leq \btau} \sbr{C_1 \Psi_1 \sbr{\tau} + C_2 \Psi_2 \sbr{\tau}},
    \end{equation}
    where $\Psi_i\sbr{\cdot}$, $i = 1, 2$, are two basis function given by
    \begin{equation}\label{eq_defPsi}
        \Psi_i \sbr{\tau} = - \bbr{\sbr{\delta_{1i} + \delta_{2i} \mathrm{I}_{\mathrm{d}}}\log f_{\varepsilon}}^{\prime \prime} \sbr{F_{\varepsilon}^{-1} \sbr{\tau}}
    \end{equation}
    with $\mathrm{I}_{\mathrm{d}}$ the identity function, i.e., $\mathrm{I}_{\mathrm{d}}\sbr{y} = y$ for $y \in \bR$;
    the coefficients $C_1$ and $C_2$ are coefficients selected to satisfy the corresponding conditions in \eqref{eq_cons_J} or \eqref{eq_cons_Jsgm}
    i.e.,
    \begin{equation*}
        \begin{aligned}
            \text{if} \;\; z &= m, \;\; C_i = \frac{\delta_{1i} A_{12} - \delta_{2i} A_{11}}{A_{01}A_{12} - A_{02}A_{11}}, \quad i = 1, 2,\\
            \text{if} \;\; z &= \sigma, \;\; C_i = \frac{\delta_{2i} A_{01} - \delta_{1i} A_{02}}{A_{01}A_{12} - A_{02}A_{11}}, \quad i = 1, 2,
        \end{aligned}
    \end{equation*}
    with
    \begin{equation*}
        A_{kj} = \int_{[\utau, \btau]} \sbr{F_{\varepsilon}^{-1} \sbr{\tau}}^k \Psi_j \sbr{\tau} \di \tau,
    \end{equation*}
    for $k=0, 1$, and $j = 1, 2$.
\end{theorem}

The following remark shows that Theorem~\ref{thm_optJ} is an extension of existing results about the optimal weight functions in L-Estimation.

\begin{remark}
    Theorem~\ref{thm_optJ} allows us to find the optimal weight function under the constraint $\mathrm{Supp}\bbr{J\sbr{\cdot}} \subset [\utau, \btau]$.
    This constraint is artificially introduced to control the error caused from LPI (see Remark~\ref{rmk_ResWeight}).
    If we consider a special case of Theorem~\ref{thm_optJ}, where $[\utau, \btau] = [0, 1]$, i.e, the constraint on $\mathrm{Supp}\bbr{J\sbr{\cdot}}$ is removed,
    then the optimal weight function $J_m^*\sbr{\cdot}$ (resp. $J_{\sigma}^*\sbr{\cdot}$) degenerates into $\Psi_1 \sbr{\cdot}$ (resp. $\Psi_2 \sbr{\cdot}$), which is just the optimal score functions for $m(x)$ (reps. $\sigma(x)$) given in \cite{Portnoy1989Adaptive, Koenker2005}.
\end{remark}

The optimal weight function in \eqref{eq_Jopt} is deeply related to the distribution function and the PDF of the error, which can be difficult to estimate especially in the case of streaming data.
In the following corollary, we manage to express $J_z^*\sbr{\cdot}$ using the CDF $F_{Y\vert x}\sbr{\cdot}$.
\begin{corollary}\label{coro_JoptFyx}
    Under Assumption~\ref{assum_fmsigma}, the optimal weight function in \eqref{eq_Jopt} can be expressed as
    \begin{equation}\label{eq_Jopttilde}
        J_z^*\sbr{\tau} = I\sbr{\utau \leq \tau \leq \btau} (\tilde{C}_1 \tilde{\Psi}_1 \sbr{\tau} + \tilde{C}_2 \tilde{\Psi}_2 \sbr{\tau}),
    \end{equation}
    where $\tilde{\Psi}_i\sbr{\cdot}$, $i = 1, 2$, are basis functions given by
    \begin{equation*}
        \tilde{\Psi}_i\sbr{\tau} = -\int_{I_*} \{\sbr{\delta_{1i} + \delta_{2i} \mathrm{I}_{\mathrm{d}}} \log F_{Y\vert x}^{\prime}\}^{\prime \prime} \sbr{F_{Y\vert x}^{-1} \sbr{\tau}} \di x
    \end{equation*}
    and the coefficients $\tilde{C}_1$ and $\tilde{C}_2$ are selected to fulfill the corresponding conditions $(\mathrm{C}_{z1}, \mathrm{C}_{z2}^{\prime})$, i.e.,
    \begin{equation*}
    \begin{aligned}
        &\text{if} \;\; z = m, \;\; \left\{\begin{aligned}
            &\tilde{C}_1 = \frac{d_{0} \tilde{A}_{12} - d_{1} \tilde{A}_{02}}{\tilde{A}_{01}\tilde{A}_{12} - \tilde{A}_{02}\tilde{A}_{11}}, \\
            &\tilde{C}_2 = \frac{-d_{0} \tilde{A}_{11} + d_{1} \tilde{A}_{01}}{\tilde{A}_{01}\tilde{A}_{12} - \tilde{A}_{02}\tilde{A}_{11}}, \end{aligned}\right. \\
        &\text{if} \;\; z = \sigma, \;\; \left\{\begin{aligned}
                &\tilde{C}_1 \tilde{A}_{01} + \tilde{C}_2 \tilde{A}_{02} = 0, \\
                &\tilde{C}_1^2 \tilde{D}_{1, 1} + 2 \tilde{C}_1 \tilde{C}_2 \tilde{D}_{1, 2} + \tilde{C}_2^2 \tilde{D}_{2, 2} = d_3^2, \end{aligned}\right.
    \end{aligned}
    \end{equation*}
    with $d_{0} =  1$, $d_1 = \E{W\sbr{X}Y}$ and
    \begin{equation*}
        d_3^2 = \E{W(X) (Y^2 - m^2 \sbr{X})},
    \end{equation*}
    \begin{equation*}
        \begin{aligned}
            \tilde{A}_{kj} =\;& \int_{I_*}\int_{\bR} \bbr{y W\sbr{x} f_{X}\sbr{x}}^k \\
            &\quad \times \tilde{\Psi}_j(F_{Y\vert x}\sbr{y}) \di F_{Y\vert x}\sbr{y}\di x,
        \end{aligned}
    \end{equation*}
    for $k = 0, 1$ and $j = 1, 2$.
\end{corollary}

Corollary~\ref{coro_JoptFyx} suggests a renewable estimation method for the optimal weight function based on the interpolated empirical CDF given in \eqref{eq_hatintpFYx}.
Specifically speaking, the estimator of $J_z^*\sbr{\cdot}$ can be given by \eqref{eq_Jopttilde} with the unknown $F_{Y\vert x}\sbr{\cdot}$ and its derivatives replaced by $\I_l \widehat{F}_{Y\vert x, t}\sbr{\cdot}$ and its corresponding derivatives, and the involved constants $\tilde{C}_i$ can be estimated by renewable procedures similar to the ones introduced in \eqref{eq_estw} and~\eqref{eq_hatthetat}.

\section{Numerical Experiments}\label{sec_simu}

In this section, we conduct simulation studies and real data analyses to verify performance of various estimators involved in this paper.

In the numerical experiments, all the involved kernel functions are taken as the Epanechnikov kernel, i.e., $K(z) = \max\bbr{0, 3/4 \left(1-z^{2}\right)}$.
To implement the renewable WCQR estimation, $3$rd-degree LPI is used to obtain the interpolated empirical CDF.
To obtain the LPI nodes, the set $G_*$ is formed by $100$ grid points evenly distributed over the intervals $I_*$.
And for each $x_i$ in $G_*$, the set $G_{x_i}$ consists of the points $y_{ij}$ given by
\begin{equation*}
    y_{ij} = \argmin_{y \in \bR} \sum_{q: \; X_q \in \mathcal{N}_k \sbr{x_i, \D_{\mathrm{val}}}} \rho_{\tau_j}\sbr{Y_q - y}
\end{equation*}
for $\tau_j = j /100$ and $j = 1, \cdots, 99$.
Here $\D_{\mathrm{val}}$ is a validation data set containing the first $2000$ samples collected from the data stream.
And the function $\mathcal{N}_k$ is defined in Section~\ref{sec_LPI} with $k$ empirically selected as $\max\bbr{0.1\#\D_{\mathrm{val}}, \#G_{x_i}}$.

For implementations, we need to artificially determine a terminal time $T$ for the streaming data sets.
Depending on whether or not the cumulative number $N_T$ is supposed to be known, we consider two bandwidth selectors:
the ``oracle'' optimal constant bandwidth $\tilde{h}_t^*$ from~\eqref{eq_optht} and the renewable constant bandwidth $\widehat{h}_t$ from \eqref{eq_subhtx}.
The unknown constant $C_h$ in \eqref{eq_optht} and \eqref{eq_subhtx} is replaced by its estimator $\widehat{C}_h$ obtained by $10$-fold Cross-Validation on the validation data set $\D_{\mathrm{val}}$.

We mainly discuss the performance of the following three proposed renewable estimators:
\begin{itemize}
    \item $\widehat{r}_{\mathrm{ntm}}$: It is the renewable NTM $\widetilde{r}_T(x; J_{m, 0.5})$ given by \eqref{eq_widehatthetaF} with the weight function $J_{m, 0.5}$ given in \eqref{eq_Jalpw} with $\alpha = 0.1$; the bandwidth is selected as the renewable $\widehat{h}_t$ given in \eqref{eq_subhtx} with $C_h$ replaced by $\widehat{C}_h$.
    \item $\widehat{r}_{\mathrm{bctm}}$: It is the renewable BCTM $\widetilde{r}_T \sbr{x; J_{m, \widehat{w}_T}}$ given by \eqref{eq_widehatthetaF} with the weight function $J_{m, \widehat{w}_T}$ given in~\eqref{eq_Jalpw} with $\alpha = 0.1$ and $\widehat{w}_T$ obtained by \eqref{eq_estw}; the bandwidth selector is the same with that of $\widehat{r}_{\mathrm{ntm}}$.
    \item $\widehat{r}_{\mathrm{rtsd}}$: It is the renewable RTSD $\widetilde{r}_T (x; J_{\sigma, \widehat{\theta}_T})$ given by \eqref{eq_widehatthetaF} with the weight function $J_{\sigma, \widehat{\theta}_T}$ given in~\eqref{eq_Jsigtheta} with $\alpha = 0.1$ and $\widehat{\theta}_T$ obtained by \eqref{eq_hatthetat}; the bandwidth selector is the same with that of $\widehat{r}_{\mathrm{ntm}}$.
\end{itemize}
For comparison, we introduce the following benchmark estimators:
\begin{itemize}
    \item $\widehat{r}_{\mathrm{ntm}}^*$, $\widehat{r}_{\mathrm{bctm}}^*$ and $\widehat{r}_{\mathrm{rtsd}}^*$: They are the oracle counterparts of $\widehat{r}_{\mathrm{ntm}}$, $\widehat{r}_{\mathrm{bctm}}$ and $\widehat{r}_{\mathrm{rtsd}}$, respectively, i.e., all of them are computed on the full data $\cup_{t\leq T}\D_t$ with the bandwidth selected as the estimated oracle optimal constant bandwidths, i.e., $h = \widehat{C}_h N_T^{-1/5}$.
    \item $\widehat{r}_{\mathrm{ntm}}^{\mathrm{a}}$, $\widehat{r}_{\mathrm{bctm}}^{\mathrm{a}}$ and $\widehat{r}_{\mathrm{rtsd}}^{\mathrm{a}}$: They are the simple-average counterparts of $\widehat{r}_{\mathrm{ntm}}$, $\widehat{r}_{\mathrm{bctm}}$ and $\widehat{r}_{\mathrm{rtsd}}$, respectively, which are obtained by simply averaging the corresponding local estimators computed on the data chunks $\D_1, \cdots, \D_T$.
    For example, $\widehat{r}_{\mathrm{ntm}}^{\mathrm{a}}\sbr{x} = 1/T \sum_{t=1}^T \widehat{r}_{\mathrm{ntm}}^{[t]}\sbr{x}$, where $\widehat{r}_{\mathrm{ntm}}^{[t]}$ is the analogue of $\widehat{r}_{\mathrm{ntm}}^*$ computed on $\D_t$.
    Here all the bandwidths are selected as the estimated local optimal constant bandwidths as $h_t = \widehat{C}_h n_t^{-1/5}$.
    \item $\widehat{r}_{\mathrm{nw}}^*$ and $\widehat{r}_{\mathrm{nwsd}}^*$: They are the oracle Nadaraya-Watson (NW) estimators for $m\sbr{x}$ and $\sigma\sbr{x}$, respectively, which are computed on the full data $\cup_{t\leq T}\D_t$, i.e.,
    \begin{equation*}
        \widehat{r}_{\mathrm{nw}}^*\sbr{x} = \frac{\sum_{t=1}^T \sum_{j=1}^{n_t} Y_{tj} K_{h}\left(X_{tj} - x\right)}{\sum_{t=1}^T \sum_{j=1}^{n_t} K_{h}\left(X_{tj} - x\right)},
    \end{equation*}
    and
    \begin{align*}
        &\sbr{\widehat{r}_{\mathrm{nwsd}}^*}^2\sbr{x} \\
        =\;& \frac{\sum_{t=1}^T \sum_{j=1}^{n_t} \sbr{Y_{tj} - \widehat{r}_{\mathrm{nw}}^*\sbr{x}}^2 K_{h}\left(X_{tj} - x\right)}{\sum_{t=1}^T \sum_{j=1}^{n_t} K_{h}\left(X_{tj} - x\right)},
    \end{align*}
    where the bandwidth is selected as the oracle optimal constant bandwidths, i.e., $h = C_{h}^{\prime} N_T^{-1/5}$ with the constant $C_{h}^{\prime}$ estimated by $10$-fold Cross-Validation on $\D_{\mathrm{val}}$.
\end{itemize}

\begin{table*}[ht]
    \centering
    \renewcommand\arraystretch{1.2}
    \caption{The performance comparison between the renewable estimators and their oracle counterparts under streaming data with various chunk sizes, the full sample size is $N_T = 10^5$, the multiplying factor of contaminated data is $\lambda = 1$.}
    \label{tab_datapart}
    \resizebox{\textwidth}{!}{%
    \begin{tabular}{@{}cclccccccccccc@{}}
        \toprule
        \multirow{2}{*}{Example} & \multirow{2}{*}{Error distribution} & \multirow{2}{*}{Indicator} & \multicolumn{2}{c}{$n_t \equiv 10^4$} &  & \multicolumn{2}{c}{$n_t \equiv 10^3$} &  & \multicolumn{2}{c}{$n_t \equiv 10^2$} &  & \multicolumn{2}{c}{$n_t \equiv 10$} \\ \cmidrule(lr){4-5} \cmidrule(lr){7-8} \cmidrule(lr){10-11} \cmidrule(l){13-14}
         &  &  & mean & std. &  & mean & std. &  & mean & std. &  & mean & std. \\ \midrule
        1a & N(0, 1) & RASE$\sbr{\widehat{r}_{\mathrm{ntm}}^*,   \widehat{r}_{\mathrm{ntm}}}$ & 0.9043 & 0.0304 &  & 0.9940 & 0.0466 &  & 1.0248 & 0.0401 &  & 0.9189 & 0.0416 \\
        &  & RASE$\sbr{\widehat{r}_{\mathrm{ntm}}^*,   \widehat{r}_{\mathrm{ntm}}^{\mathrm{a}}}$ & 0.8989 & 0.0498 &  & 0.1525 & 0.0342 &  & 0.0007 & 0.0000 &  & -- & -- \\
         &  & RASE$\sbr{\widehat{r}_{\mathrm{bctm}}^*,   \widehat{r}_{\mathrm{bctm}}}$ & 0.8901 & 0.0218 &  & 0.9404 & 0.0477 &  & 1.0069 & 0.0416 &  & 0.9225 & 0.0420 \\
         &  & RASE$\sbr{\widehat{r}_{\mathrm{bctm}}^*,   \widehat{r}_{\mathrm{bctm}}^{\mathrm{a}}}$ & 0.7856 & 0.0253 &  & 0.4210 & 0.0176 &  & 0.0017 & 0.0009 &  & -- & -- \\ \\
        1b & Pareto(3) & RASE$\sbr{\widehat{r}_{\mathrm{bctm}}^*,   \widehat{r}_{\mathrm{bctm}}}$ & 1.0606 & 0.0401 &  & 1.0257 & 0.0433 &  & 1.0307 & 0.0473 &  & 1.0216 & 0.0421 \\
        &  & RASE$\sbr{\widehat{r}_{\mathrm{bctm}}^*,   \widehat{r}_{\mathrm{bctm}}^{\mathrm{a}}}$ & 0.6277 & 0.0548 &  & 0.4033 & 0.0731 &  & 0.0016 & 0.0004 &  & -- & -- \\
         &  & RASE$\sbr{\widehat{r}_{\mathrm{rtsd}}^*,   \widehat{r}_{\mathrm{rtsd}}}$ & 1.0002 & 0.0004 &  & 1.0005 & 0.0004 &  & 1.0004 & 0.0004 &  & 1.0004 & 0.0004 \\
         &  & RASE$\sbr{\widehat{r}_{\mathrm{rtsd}}^*,   \widehat{r}_{\mathrm{rtsd}}^{\mathrm{a}}}$ & 0.5221 & 0.0471 &  & 0.1362 & 0.0288 &  & 0.0013 & 0.0012 &  & -- & -- \\ \bottomrule
    \end{tabular}%
    }
\end{table*}

\subsection{Simulation Studies}

In the simulation studies, we will consider various experiment conditions, such as homoscedastic or heteroscedastic models, and symmetric or asymmetric errors.
To this end, we consider the following two models:
\begin{align*}
    \mathrm{Model 1: } &Y = \sin (2 X)+2 \exp \left(-16 X^{2}\right) + 0.5 \varepsilon \\
    &\sptext{with} X \sim N(0, 1), \quad I_* = [-1.5, 1.5],\\
    \mathrm{Model 2: } &Y = X \sin (2 \pi X) + \left(2 + \cos(2\pi X)\right) \varepsilon \\
    &\sptext{with} X \sim U(0, 1), \quad I_* = [0, 1],
\end{align*}
where Model 1 is homoscedastic and adopted from \cite{Fan1992Variable}, and Model 2 is heteroscedastic and adopted from \cite{Kai2010}
We consider various kinds of distributions of $\varepsilon$.
Moreover, we also use the mixtures of two error distributions to model so-called contaminated data.
Specifically, a mixture distribution is chosen as $0.95 F_{\varepsilon} + 0.05 F_{\lambda\,\varepsilon}$ with a multiplying factor $\lambda$, where $F_{\varepsilon}$ is the distribution function of $\varepsilon$.
If without special statement, all the distributions of $\varepsilon$ involved in the simulations are centralized.
Based on the combinations of different models and error distributions, we consider the following four examples:
\begin{itemize}
    \item Example 1a: the Model 1 with various symmetric error distributions;
    \item Example 1b: the Model 1 with various asymmetric error distributions;
    \item Example 2a: the Model 2 with various symmetric error distributions;
    \item Example 2b: the Model 2 with various asymmetric error distributions.
\end{itemize}
To model the streaming data, we generate the full data of size $N_T$ from the considered models and equally divide the full data into $T$ data chunks.

\begin{table*}[ht]
    \centering
    \caption{The performance comparison between the oracle NW estimators and the renewable WCQR estimators under various models with contaminated streaming data, the full sample size is $N_T = 10^5$ and the chunk size is $n_t \equiv 1000$.}
    \label{tab_data_corru}
    \resizebox{0.8\textwidth}{!}{%
    \begin{tabular}{@{}ccccccccccc@{}}
        \toprule
        \multirow{2}{*}{Example} & \multirow{2}{*}{Error distribution} & \multirow{2}{*}{$\lambda$} & \multicolumn{2}{c}{RASE$\sbr{\widehat{r}_{\mathrm{nw}}^*,   \widehat{r}_{\text{ntm}}}$} &  & \multicolumn{2}{c}{RASE$\sbr{\widehat{r}_{\mathrm{nw}}^*,   \widehat{r}_{\text{bctm}}}$} &  & \multicolumn{2}{c}{RASE$\sbr{\widehat{r}_{\mathrm{nwsd}}^*,   \widehat{r}_{\text{rtsd}}}$} \\ \cmidrule(lr){4-5} \cmidrule(lr){7-8} \cmidrule(l){10-11}
         &  &  & mean & std. &  & mean & std. &  & mean & std. \\ \midrule
        1a & N(0, 1) & 1 & 0.9558 & 0.0528 &  & 0.9527 & 0.0512 &  & 0.7368 & 0.0765 \\
         &  & 3 & 1.1806 & 0.1248 &  & 1.1830 & 0.1231 &  & 0.9939 & 0.0352 \\
         &  & 5 & 1.6637 & 0.2761 &  & 1.6278 & 0.2609 &  & 1.0059 & 0.0227 \\
         &  & 10 & 2.7210 & 0.5874 &  & 3.1802 & 0.6356 &  & 1.0139 & 0.0158 \\
         &  &  &  &  &  &  &  &  &  &  \\
        1b & F(10, 6) & 1 & 0.0182 & 0.0048 &  & 1.1897 & 0.3326 &  & 1.1172 & 0.1190 \\
         &  & 3 & 0.0219 & 0.0072 &  & 1.5075 & 0.4427 &  & 1.0719 & 0.0994 \\
         &  & 5 & 0.0289 & 0.0089 &  & 1.7405 & 0.4581 &  & 1.0700 & 0.1068 \\
         &  & 10 & 0.0653 & 0.0199 &  & 4.1594 & 1.2333 &  & 1.0752 & 0.0829 \\
         &  &  &  &  &  &  &  &  &  &  \\
        2a & N(0, 1) & 1 & 0.8586 & 0.1025 &  & 0.8786 & 0.1127 &  & 0.8108 & 0.1005 \\
         &  & 3 & 1.1278 & 0.0948 &  & 1.1822 & 0.0922 &  & 0.9417 & 0.0182 \\
         &  & 5 & 1.3174 & 0.1746 &  & 1.8534 & 0.2941 &  & 1.0508 & 0.0135 \\
         &  & 10 & 3.2014 & 0.7741 &  & 2.6362 & 0.6085 &  & 1.0620 & 0.0067 \\
         &  &  &  &  &  &  &  &  &  &  \\
        2b & F(10, 6) & 1 & 0.0237 & 0.0072 &  & 1.1536 & 0.2070 &  & 0.9569 & 0.0395 \\
         &  & 3 & 0.0279 & 0.0072 &  & 1.1630 & 0.2657 &  & 0.9941 & 0.0354 \\
         &  & 5 & 0.0389 & 0.0182 &  & 1.6001 & 0.6350 &  & 1.0865 & 0.0362 \\
         &  & 10 & 0.0683 & 0.0238 &  & 2.2746 & 0.7554 &  & 1.0768 & 0.0231 \\ \bottomrule
    \end{tabular}%
    }
\end{table*}

\begin{table*}[tb]
    \centering
    \caption{The performance comparison between the oracle NW estimators and the renewable WCQR estimators under various models and error distributions, the full sample size is $N_T = 10^5$, the chunk size is $n_t \equiv 1000$ and the multiplying factor of contaminated data is $\lambda = 1$.}
    \label{tab_error_distribution}
    \resizebox{0.8\textwidth}{!}{%
    \begin{tabular}{@{}clcccccccc@{}}
    \toprule
    \multirow{2}{*}{Example} & \multirow{2}{*}{Error distribution} & \multicolumn{2}{c}{RASE$\sbr{\widehat{r}_{\mathrm{nw}}^*,   \widehat{r}_{\text{ntm}}}$} &  & \multicolumn{2}{c}{RASE$\sbr{\widehat{r}_{\mathrm{nw}}^*,   \widehat{r}_{\text{bctm}}}$} &  & \multicolumn{2}{c}{RASE$\sbr{\widehat{r}_{\mathrm{nwsd}}^*,   \widehat{r}_{\text{rtsd}}}$} \\ \cmidrule(lr){3-4} \cmidrule(lr){6-7} \cmidrule(l){9-10}
     &  & mean & std. &  & mean & std. &  & mean & std. \\ \midrule
    1a & Standard Laplace & 1.3054 & 0.1579 &  & 1.2949 & 0.1558 &  & 0.9860 & 0.0145 \\
     & t(3) & 1.5500 & 0.2860 &  & 1.5221 & 0.2757 &  & 1.0369 & 0.0801 \\
     &  &  &  &  &  &  &  &  &  \\
    1b & F(4, 6) & 0.0114 & 0.0036 &  & 1.2578 & 0.3805 &  & 1.0196 & 0.0720 \\
     & Pareto(3) & 0.0151 & 0.0063 &  & 1.2186 & 0.3968 &  & 1.0700 & 0.1414 \\
     &  &  &  &  &  &  &  &  &  \\
    2a & Standard Laplace & 1.3050 & 0.1538 &  & 1.2607 & 0.1442 &  & 0.9280 & 0.0081 \\
     & t(3) & 1.6960 & 0.2882 &  & 1.6480 & 0.2751 &  & 0.9598 & 0.0331 \\
     &  &  &  &  &  &  &  &  &  \\
    2b & F(10, 6) & 0.0237 & 0.0072 &  & 1.1536 & 0.2070 &  & 0.9569 & 0.0395 \\
     & Lognorm(0, 1) & 0.0137 & 0.0036 &  & 1.1069 & 0.2369 &  & 0.9527 & 0.0092 \\ \bottomrule
    \end{tabular}%
    }
\end{table*}

For each estimation, the number of replications in the simulation is designed as $200$.
The performance of any estimator $\widehat{g}(\cdot)$ of a function $g(\cdot)$ is evaluated by the average squared errors (ASEs) defined by
$$\operatorname{ASE}(\widehat{g})= \frac{1}{\#G_*} \sum_{x_i \in G_*} \left\vert \widehat{g}\left(x_i\right)-g\left(x_i\right)\right\vert ^{2}.$$
To compare the performance of two estimators $\widehat{g}_1$ and $\widehat{g}_2$, we use the ratio of average squared errors (RASEs):
$$\mathrm{RASE}\sbr{\widehat{g}_1,\widehat{g}_2} = \frac{\operatorname{ASE}(\widehat{g}_1)}{\operatorname{ASE}(\widehat{g}_2)}$$

\subsubsection{Scenario 1: Streaming Data with varying Chunk Size}

We first discuss the influence of data partitioning on our WCQR estimators.
To this end, we fix the full sample size $N_T = 10^5$ and successively take the chunk number $T = 10, 10^2, 10^3, 10^4$, resulting the chunk sizes $n_t = 10^4, 10^3, 10^2, 10$, respectively.
The multiplying factor is $\lambda = 1$, i.e., the streaming data is not contaminated.
Then we test the WCQR estimators in Example 1a and 1b with the associated simple-average estimators and oracle estimators used as benchmarks.
The relevant RASEs are reported in Table~\ref{tab_datapart}.

From Table~\ref{tab_datapart}, we can see that the renewable WCQR estimators $\widehat{r}_{\mathrm{ntm}}$, $\widehat{r}_{\mathrm{bctm}}$ and $\widehat{r}_{\mathrm{rtsd}}$ perform well with all the associated RASEs not only insensitive to chunk size $n_t$ but also closed to $1$.
On the contrary, the simple-average estimators $\widehat{r}_{\mathrm{ntm}}^{\mathrm{a}}$, $\widehat{r}_{\mathrm{bctm}}^{\mathrm{a}}$ and $\widehat{r}_{\mathrm{rtsd}}^{\mathrm{a}}$ is susceptible to the chunk size.
For small chunk sizes, i.e., $n_t \leq 10^3$, their performances are significantly inferior than that of the oracle estimators.
In the extreme case $n_t = 10$, the renewable WCQR estimators still work well, but the simple-average estimators can not give results because the data chunk is too small to compute the local estimators.
These imply that our renewable algorithm enjoys desirable performance robust to the chunk size of the streaming data,
and the obtained renewable WCQR estimators are comparable to the oracle ones.

\subsubsection{Scenario 2: Models with Contaminated Streaming Data}

We turn to test the robustness and model adaptiveness of our WCQR estimators. The benchmark estimators are the oracle NW estimators.
We successively take various multiplying factors $\lambda = 1, 3, 5, 10$ for contaminated streaming data.
Since all the involved estimators are impervious to the data partitioning, we only consider a fixed chunk size $n_t=100$ with the full data size $N_T = 10^5$.
Then we test the estimators in the four examples and report the relevant RASEs in Table~\ref{tab_data_corru}.

For the results in Table~\ref{tab_data_corru}, we have the following discussions:
\begin{itemize}
    \item[(i)] In Examples 1a and 2a, the model is symmetric and the two estimators $\widehat{r}_{\mathrm{ntm}}$ and $\widehat{r}_{\mathrm{bctm}}$  show similar behaviors.
    Specifically, when $\gamma = 1$, i.e., the error is normal without contaminations, both of them are slightly inferior than the oracle NW estimator $\widehat{r}_{\mathrm{nw}}^*$.
    However, when $\lambda > 1$, i.e., the streaming data are contaminated, both of them outperform $\widehat{r}_{\mathrm{nw}}^*$.
    Moreover, the values of $\mathrm{RASE}\sbr{\widehat{r}_{\mathrm{nw}}^*, \widehat{r}_{\text{ntm}}}$ and $\mathrm{RASE}\sbr{\widehat{r}_{\mathrm{nw}}^*, \widehat{r}_{\text{bctm}}}$ increase as $\lambda$ is increasing, which suggests that compare with the NW estimator, the NTM and BCTM are more robust to data contaminations.
    We also notice that even if $\lambda$ is large, there is no obvious gap between the performance of $\widehat{r}_{\text{ntm}}$ and $\widehat{r}_{\text{bctm}}$.
    This justifies our claim in Remark~\ref{rmk_EWY} that the robustness of the BCTM is not susceptible to the non-robust estimation of $E_{WY}$.
    \item[(iii)] We focus on the results of $\widehat{r}_{\mathrm{ntm}}$ and $\widehat{r}_{\mathrm{bctm}}$ in Examples 1b and 2b, where the model is asymmetric.
    Contrary to the case of symmetric models, the behaviors between the NTM and the BCTM are quite different.
    All the values of $\mathrm{RASE}\sbr{\widehat{r}_{\mathrm{nw}}^*, \widehat{r}_{\mathrm{bctm}}}$ are closed to zero, indicating that the NTM is far inferior to the NW estimator and it can be inconsistent for asymmetric models.
    While, the values of $\mathrm{RASE}\sbr{\widehat{r}_{\mathrm{nw}}^*, \widehat{r}_{\mathrm{bctm}}}$ suggest that the BCTM still works well and even outperforms the NW estimators when the data is contaminated.
    The above results show that under the weight selection criterions in Section~\ref{sec_modelweighsel}, our renewable WCQR estimator is adaptive to symmetric and asymmetric models.
    \item[(iv)] We turn to discuss the results of the RTSD $\widehat{r}_{\text{rtsd}}$ in the four examples.
    As $\lambda$ is increasing, the RASEs between $\widehat{r}_{\text{nwsd}}$ and $\widehat{r}_{\text{rtsd}}$ show an increasing trend with the values uniformly larger than $1$ when $\lambda > 3$.
    This means that our estimator $\widehat{r}_{\text{rtsd}}$ is more robust than $\widehat{r}_{\text{nwsd}}$ and enjoy more advantages when the streaming data are contaminated.
\end{itemize}

\subsubsection{Scenario 3: Models with Nonnormal Error Distributions}

We focus on the performance of our WCQR estimators for non-normal error distributions.
We still consider a fixed chunk size $n_t=100$ with the full data size $N_T = 10^5$.
The multiplying factor is $\lambda = 1$, i.e., there is no contaminated data.
We use the the NW estimators as a benchmark, and test the WCQR estimators in the four examples with nonnormal error distributions.
The relevant RASEs are reported in Table~\ref{tab_error_distribution}.

From Table~\ref{tab_error_distribution}, we have the following finds:
\begin{itemize}
    \item[(i)] For symmetric models, i.e. Examples 1a and 2a, both of the NTM $\widehat{r}_{\mathrm{ntm}}$ and the BCTM $\widehat{r}_{\mathrm{bctm}}$ show advantage over the NW estimator $\widehat{r}_{\mathrm{nw}}^*$, which suggests that our WCQR estimators can be more efficient than the NW estimators when the error is nonnormal. The phenomenon is in line with the feature of CQR method in \citep{Zou2008Composite, Kai2010}.
    \item[(ii)] For asymmetric models, i.e., Examples 1b and 2b, benefit from the model adaptiveness mentioned in the last scenario, $\widehat{r}_{\mathrm{bctm}}$ maintains its advantage over $\widehat{r}_{\mathrm{nw}}^*$. And as expected, the NTM $\widehat{r}_{\mathrm{ntm}}$ does not work because of the nonnegligible bias arising in asymmetric models.
    \item[(iii)] The results of $\mathrm{RASE}\sbr{\widehat{r}_{\mathrm{nwsd}}^*, \widehat{r}_{\text{rtsd}}}$ show that the RTSD $\widehat{r}_{\text{rtsd}}$ seem not superior than $\widehat{r}_{\mathrm{nwsd}}^*$. This is not surprising, because the weight function $J_{\sigma, \widehat{\theta}_T}$ used by the RTSD is generally not optimal in terms of estimation variance. Moreover, we also note that $\widehat{r}_{\text{rtsd}}$ is a  renewable estimator obtained from streaming data, but $\widehat{r}_{\mathrm{nwsd}}^*$ is an oracle estimator directly computed on the full data set.
\end{itemize}

\subsection{Real Data Example}

For case study, we apply our method to the Beijing Multi-Site Air-Quality Data set from the UCI machine learning repository
\footnote[1]{\url{https://archive-beta.ics.uci.edu/ml/datasets/beijing+multi+site+air+quality+data}}.
This data set consists of hourly data about 6 main air pollutants and 6 relevant meteorological variables collected from 12 nationally-controlled air-quality monitoring sites in Beijing, China.
The observational data cover the time period from March 1st, 2013 to February 28th, 2017, and the 420768 observed values of each variable.
Our goal is to fit the relationship between the main air pollutants and the relevant meteorological variables in the dataset.

From the data set, we select two pairs of air pollutants and meteorological variables as the response variable $Y$ and the covariate $X$ in model~\eqref{eq_model}, and then we obtain two examples listed in Table~\ref{tab_realexp}.
In Table~\ref{tab_realexp}, the terms DEWP, O3, WSPM and PM10 are abbreviations to dew point temperature (degree Celsius), O3 concentration (ug/m\^3), wind speed (m/s) and PM10 concentration (ug/m\^3), respectively.

\begin{table}[t]
    \centering
    \caption{The relations studied in Real Data Example}
    \label{tab_realexp}
    \begin{tabular}{@{}ccccc@{}}
    \toprule
    Example & X & Y & \#$\D_{\mathrm{train}}$ & \#$\D_{\mathrm{test}}$ \\ \midrule
    3a & DEWP & O3 & 305572 & 101808 \\
    3b & WSPM & PM10 & 311146 & 103146 \\ \bottomrule
    \end{tabular}%
\end{table}

\begin{table*}[ht]
    \centering
    \caption{The RMSEs of various estimators in fitting the test set from the Beijing Multi-Site Air-Quality Data set}
    \label{tab_realdata_rmse}
    \resizebox{0.8\textwidth}{!}{%
    \begin{tabular}{@{}ccccccccccccc@{}}
    \toprule
    Example & $\gamma$ & Data chunk levels & $\widehat{r}_{\text{ntm}}^*$ & $\widehat{r}_{\text{ntm}}$ & $\widehat{r}_{\text{ntm}}^{\text{a}}$ &  & $\widehat{r}_{\text{bctm}}^*$ & $\widehat{r}_{\text{bctm}}$ & $\widehat{r}_{\text{bctm}}^{\text{a}}$ &  & $\widehat{r}_{\mathrm{nw}}^*$ \\ \midrule
    3a & 0 & Monthly & 53.168 & 53.163 & 58.669 &  & 52.747 & 52.740 & 60.817 &  & 52.694 \\
     &  & Daily &  & 53.197 & 65.282 &  &  & 52.779 & 63.218 &  &  \\
     &  & Hourly &  & 53.165 & 66.347 &  &  & 52.751 & 63.286 &  & \\
     &  &  &  &  &  &  &  &  &  &  &  &  \\
     & 200 & Monthly & 53.018 & 53.004 & 59.257 &  & 53.211 & 53.196 & 61.913 &  & 55.210 \\
     &  & Daily &  & 53.063 & 65.395 &  &  & 53.257 & 61.279 &  &  \\
     &  & Hourly &  & 52.919 & 66.042 &  &  & 52.541 & 64.832 &  &  \\
     &  &  &  &  &  &  &  &  &  &  &  &  \\
     & 300 & Monthly & 53.145 & 53.130 & 60.294 &  & 53.892 & 53.874 & 62.241 &  & 54.038 \\
     &  & Daily &  & 53.017 & 66.648 &  &  & 53.889 & 66.288 &  &  \\
     &  & Hourly &  & 52.928 & 65.717 &  &  & 53.880 & 59.579 &  &  \\
     &  &  &  &  &  &  &  &  &  &  &  &  \\
     & 500 & Monthly & 52.974 & 52.960 & 62.377 &  & 55.559 & 55.545 & 65.395 &  & 57.476 \\
     &  & Daily &  & 52.931 & 66.749 &  &  & 55.507 & 63.527 &  &  \\
     &  & Hourly &  & 52.835 & 67.130 &  &  & 55.535 & 68.755 &  &  \\
     &  &  &  &  &  &  &  &  &  &  & \\
     & 800 & Monthly & 53.003 & 52.990 & 65.289 &  & 59.933 & 59.920 & 61.698 &  & 60.573 \\
     &  & Daily &  & 53.024 & 70.791 &  &  & 59.980 & 71.584 &  &  \\
     &  & Hourly &  & 52.984 & 69.802 &  &  & 59.143 & 65.848 &  &  \\
     &  &  &  &  &  &  &  &  &  &  &  \\
    3b & 0 & Monthly & 91.956 & 91.963 & 95.182 &  & 91.346 & 91.350 & 95.737 &  & 91.412 \\
     &  & Daily &  & 91.987 & 106.143 &  &  & 91.382 & 106.577 &  &  \\
     &  & Hourly &  & 91.956 & 126.516 &  &  & 91.377 & 116.189 &  &  \\
     &  &  &  &  &  &  &  &  &  &  &  \\
     & 200 & Monthly & 91.825 & 91.818 & 95.501 &  & 91.839 & 91.819 & 97.605 &  & 93.098 \\
     &  & Daily &  & 91.846 & 106.158 &  &  & 91.913 & 100.807 &  &  \\
     &  & Hourly &  & 91.826 & 125.766 &  &  & 100.772 & 166.941 &  &  \\
     &  &  &  &  &  &  &  &  &  &  &  \\
     & 300 & Monthly & 91.847 & 91.838 & 94.179 &  & 92.560 & 92.657 & 98.384 &  & 94.072 \\
     &  & Daily &  & 91.832 & 158.011 &  &  & 92.571 & 141.565 &  &  \\
     &  & Hourly &  & 91.842 & 172.405 &  &  & 92.165 & 159.528 &  &  \\
     &  &  &  &  &  &  &  &  &  &  & \\
     & 500 & Monthly & 91.823 & 91.825 & 96.905 &  & 94.276 & 94.289 & 95.005 &  & 103.734 \\
     &  & Daily &  & 91.831 & 153.407 &  &  & 94.559 & 143.773 &  &  \\
     &  & Hourly &  & 91.831 & 160.089 &  &  & 92.859 & 152.451 &  &  \\
     &  &  &  &  &  &  &  &  &  &  &  \\
     & 800 & Monthly & 91.775 & 91.778 & 98.658 &  & 99.216 & 99.162 & 100.999 &  & 106.553 \\
     &  & Daily &  & 91.863 & 162.436 &  &  & 99.334 & 153.601 &  &  \\
     &  & Hourly &  & 91.835 & 185.107 &  &  & 99.321 & 166.387 &  &  \\ \bottomrule
    \end{tabular}%
    }
\end{table*}

\begin{table*}[ht]
    \centering
    \caption{The MAEs of various estimators in fitting the data in the test set from the Beijing Multi-Site Air-Quality Data set}
    \label{tab_realdata_mae}
    \resizebox{0.8\textwidth}{!}{%
    \begin{tabular}{@{}cccccccccccc@{}}
    \toprule
    Example & $\gamma$ & Data chunk levels & $\widehat{r}_{\text{ntm}}^*$ & $\widehat{r}_{\text{ntm}}$ & $\widehat{r}_{\text{ntm}}^{\text{a}}$ &  & $\widehat{r}_{\text{bctm}}^*$ & $\widehat{r}_{\text{bctm}}$ & $\widehat{r}_{\text{bctm}}^{\text{a}}$ &  & $\widehat{r}_{\mathrm{nw}}^*$ \\ \midrule
    3a & 0 & Monthly & 39.721 & 39.720 & 42.291 &  & 40.397 & 40.393 & 39.931 &  & 40.515 \\
     &  & Daily &  & 39.735 & 45.079 &  &  & 40.404 & 40.979 &  &  \\
     &  & Hourly &  & 39.725 & 45.711 &  &  & 40.409 & 40.294 &  &  \\
     &  &  &  &  &  &  &  &  &  &  & \\
     & 200 & Monthly & 39.791 & 39.782 & 42.368 &  & 39.669 & 39.660 & 42.149 &  & 41.601  \\
     &  & Daily &  & 39.811 & 45.264 &  &  & 39.654 & 40.132 &  &  \\
     &  & Hourly &  & 39.719 & 45.662 &  &  & 39.628 & 39.970 &  &   \\
     &  &  &  &  &  &  &  &  &  &  &   \\
     & 300 & Monthly & 39.917 & 39.907 & 43.394 &  & 39.638 & 39.628 & 42.826 &  & 40.059  \\
     &  & Daily &  & 39.791 & 45.927 &  &  & 39.534 & 42.064 &  &  \\
     &  & Hourly &  & 39.718 & 45.382 &  &  & 39.470 & 42.296 &  &   \\
     &  &  &  &  &  &  &  &  &  &  &  \\
     & 500 & Monthly & 39.762 & 39.753 & 44.338 &  & 39.621 & 39.612 & 40.789 &  & 41.603  \\
     &  & Daily &  & 39.746 & 63.445 &  &  & 39.624 & 52.976 &  &  \\
     &  & Hourly &  & 39.668 & 56.339 &  &  & 39.582 & 48.944 &  &  \\
     &  &  &  &  &  &  &  &  &  &  & \\
     & 800 & Monthly & 39.783 & 39.775 & 50.630 &  & 41.116 & 41.107 & 47.656 &  & 43.676 \\
     &  & Daily &  & 39.810 & 53.428 &  &  & 41.167 & 47.270 &  &  \\
     &  & Hourly &  & 39.782 & 56.896 &  &  & 40.822 & 47.200 &  &   \\
     &  &  &  &  &  &  &  &  &  &  &  \\
    3b & 0 & Monthly & 63.040 & 63.024 & 73.760 &  & 65.046 & 65.123 & 71.739 &  & 65.654  \\
     &  & Daily &  & 63.135 & 88.917 &  &  & 65.045 & 79.515 &  &  \\
     &  & Hourly &  & 63.072 & 111.921 &  &  & 65.098 & 93.365 &  &  \\
     &  &  &  &  &  &  &  &  &  &  &   \\
     & 200 & Monthly & 63.281 & 63.295 & 74.404 &  & 63.243 & 63.273 & 73.788 &  & 64.896  \\
     &  & Daily &  & 63.319 & 88.801 &  &  & 63.257 & 80.436 &  &  \\
     &  & Hourly &  & 63.296 & 111.117 &  &  & 82.052 & 97.375 &  &   \\
     &  &  &  &  &  &  &  &  &  &  &  \\
     & 300 & Monthly & 63.319 & 63.334 & 71.934 &  & 62.643 & 62.540 & 73.134 &  & 64.627  \\
     &  & Daily &  & 63.307 & 86.940 &  &  & 62.544 & 77.583 &  &   \\
     &  & Hourly &  & 63.326 & 107.375 &  &  & 63.419 & 92.971 &  &   \\
     &  &  &  &  &  &  &  &  &  &  &   \\
     & 500 & Monthly & 63.293 & 63.305 & 76.692 &  & 62.139 & 62.139 & 71.840 &  & 71.727  \\
     &  & Daily &  & 63.287 & 81.489 &  &  & 62.164 & 72.000 &  & \\
     &  & Hourly &  & 63.351 & 99.660 &  &  & 66.735 & 85.378 &  &  \\
     &  &  &  &  &  &  &  &  &  &  &  \\
     & 800 & Monthly & 63.234 & 63.225 & 107.322 &  & 63.344 & 63.321 & 97.311 &  & 68.131  \\
     &  & Daily &  & 63.297 & 115.269 &  &  & 63.427 & 99.265 &  &   \\
     &  & Hourly &  & 63.341 & 103.093 &  &  & 63.445 & 91.261 &  &   \\ \bottomrule
    \end{tabular}%
    }
\end{table*}

To test the performance of the involved estimators, we drop the data that suffer from data missing and then divide the remainder data set into training set $\D_{\mathrm{train}}$ and testing set $\D_{\mathrm{test}}$.
The set $\D_{\mathrm{train}}$ consists of the data collected before March 1st, 2017, which are used to obtain the involved estimators.
And the set $\D_{\mathrm{test}}$ consists of the data collected after March 1st, 2017, which are used to test the involved estimators.
The number of samples in $\D_{\mathrm{train}}$ and $\D_{\mathrm{test}}$ are listed in Table~\ref{tab_realdata_rmse}.
To model the data stream, the data in $\D_{\mathrm{train}}$ are revealed to the algorithm chronologically by chunks.
According to the reality, we consider three different sizes of data chunks, where the data chunks are respectively formed by monthly, daily and hourly data in $\D_{\mathrm{train}}$.
To simulate the data contamination, we randomly select $5$\% samples $\sbr{X_{ti}, Y_{ti}}$ from $\D_{\mathrm{train}}$ and replace by $\sbr{X_{ti}, Y_{ti} + \eta_{ti}}$, where $\eta_{ti}$ are random numbers sampled from N$\sbr{0, \gamma^2 \widehat{\sigma}^2}$ with $\widehat{\sigma}$ the sample standard deviation of all $Y_{ti}$ in $\D_{\mathrm{train}}$.

For an estimator $\widehat{g}(\cdot)$, its prediction accuracy is described by the root mean square error (RMSE) and the mean absolute error (MAE) on $\D_{\mathrm{test}}$, namely
\begin{equation*}
\begin{aligned}
    &\mathrm{RMSE}(\widehat{g}) = \sqrt{\frac{1}{n_{\mathrm{test}}} \sum_{(X_i, Y_i) \in \D_{\mathrm{test}}}\sbr{Y_i - \widehat{g}\sbr{X_i}}^2},  \\
    &\mathrm{MAE}(\widehat{g}) = \frac{1}{n_{\mathrm{test}}} \sum_{(X_i, Y_i) \in \D_{\mathrm{test}}}\abs{Y_i - \widehat{g}\sbr{X_i}},
\end{aligned}
\end{equation*}
where $n_{\mathrm{test}}$ is the number of observations in $\D_{\mathrm{test}}$.

The RMSEs and MAEs of the involved estimators are reported in Tables~\ref{tab_realdata_rmse} and \ref{tab_realdata_mae}, respectively.
From Tables~\ref{tab_realdata_rmse} and \ref{tab_realdata_mae}, we have the following findings:
\begin{itemize}
    \item[(i)] In most cases, the renewable estimators $\widehat{r}_{\mathrm{ntm}}$ and $\widehat{r}_{\mathrm{bctm}}$ show performance impervious to the data partitioning.
    Specifically, their RMSE and MAE are insensitive to the data chunk levels and are almost the same with that of the oracle estimators.
    On the contrary, the simple-average estimators $\widehat{r}_{\mathrm{bctm}}^{\mathrm{a}}$ is susceptible to the data chunk levels, and their performance deteriorates significantly when the data chunks becomes smaller.
    This shows that our renewable WCQR estimation can overcome the challenge arising from the data partitioning, and our renewable estimators enjoy asymptotic properties comparable to that of the oracle estimator obtained on the full data set.
    \item[(ii)] In all cases, the errors of the renewable NTM $\widehat{r}_{\mathrm{ntm}}$ is insensitive to $\gamma$.
    For most cases with $\gamma > 0$, i.e., the data is contaminated,  $\widehat{r}_{\mathrm{ntm}}$ is superior than $\widehat{r}_{\mathrm{nw}}^*$ in both terms of RMSE and MAE.
    Moreover, this advantage is enlarged when $\gamma$ is relatively large.
    In general, the above results suggest that our renewable WCQR estimation can achieve robustness for contaminated streaming data.
    \item[(iii)] The behavior of the renewable BCTM $\widehat{r}_{\mathrm{bctm}}$ is between that of $\widehat{r}_{\mathrm{ntm}}$ and $\widehat{r}_{\mathrm{nw}}^*$.
    Specifically speaking, when $\gamma = 0$, i.e., there is no contaminated data, $\widehat{r}_{\mathrm{bctm}}$ provides RMSE and MAE  quite close to that of $\widehat{r}_{\mathrm{nw}}^*$, and both of them are superior than the NTM in terms of RMSE.
    While, as $\gamma$ is increasing, the RMSE of $\widehat{r}_{\mathrm{bctm}}$ is enlarged but still less than the one of $\widehat{r}_{\mathrm{nw}}^*$, and moreover, its MAE does not increase significantly.
    The above results confirm our claim in Remark~\ref{rmk_EWY} that our renewable BCTM can consistently estimate the conditional mean and  enjoy robustness to some extent.
\end{itemize}

In summary, by comprehensively investigating the numerical results in various experiment conditions,
we can conclude the desirable performance of our estimation method and algorithms.

\backmatter

\bmhead{Supplementary information}

The supplementary material contains a detailed algorithm for renewable WCQR estimation, the relevant lemmas and technical proofs for the theoretical results.

\section*{Declarations}

\bmhead{Funding}
The research was supported by National Key R\&D Program of China (2018YFA0703900) and NNSF project of China (11971265).

\bmhead{Competing Interests}
The authors have no competing interests to declare that are relevant to the content of this article.


\begin{thebibliography}{}
\providecommand{\doi}[1]{\url{https://doi.org/#1}}
\bibcommenthead

\end{thebibliography}


\begin{thebibliography}{}
  \providecommand{\doi}[1]{\url{https://doi.org/#1}}
  \bibcommenthead

  \bibitem[\protect\citeauthoryear{Ashfahani and Pratama}{Ashfahani and
    Pratama}{2019}]{Andri2019Autonomous}
  Ashfahani, A. and M.~Pratama 2019.
  \newblock {\em Autonomous Deep Learning: Continual Learning Approach for
    Dynamic Environments}, pp.\  666--674.

  \bibitem[\protect\citeauthoryear{Bednar and Watt}{Bednar and
    Watt}{1984}]{Bednar1984Alpha}
  Bednar, J. and T.~Watt. 1984.
  \newblock Alpha-trimmed means and their relationship to median filters.
  \newblock {\em IEEE Transactions on Acoustics, Speech, and Signal
    Processing\/}~{\em 32\/}(1): 145--153.
  \newblock \doi{10.1109/TASSP.1984.1164279} .

  \bibitem[\protect\citeauthoryear{Bickel and Lehmann}{Bickel and
    Lehmann}{1975}]{Bickel1975Descriptive}
  Bickel, P.J. and E.L. Lehmann. 1975.
  \newblock Descriptive statistics for nonparametric models. {II}. {L}ocation.
  \newblock {\em Ann. Statist.\/}~{\em 3\/}(5): 1045--1069 .

  \bibitem[\protect\citeauthoryear{Boente and Fraiman}{Boente and
    Fraiman}{1994}]{Boente1994Local}
  Boente, G. and R.~Fraiman. 1994.
  \newblock Local {$L$}-estimators for nonparametric regression under dependence.
  \newblock {\em J. Nonparametr. Statist.\/}~{\em 4\/}(1): 91--101.
  \newblock \doi{10.1080/10485259408832603} .

  \bibitem[\protect\citeauthoryear{Bucak and Gunsel}{Bucak and
    Gunsel}{2009}]{Serhat2009Incremental}
  Bucak, S.S. and B.~Gunsel. 2009.
  \newblock Incremental subspace learning via non-negative matrix factorization.
  \newblock {\em Pattern Recognition\/}~{\em 42\/}(5): 788--797.
  \newblock \doi{https://doi.org/10.1016/j.patcog.2008.09.002} .

  \bibitem[\protect\citeauthoryear{Burden, Faires, and Burden}{Burden
    et~al.}{2015}]{burden2015numerical}
  Burden, R.L., J.D. Faires, and A.M. Burden. 2015.
  \newblock {\em Numerical analysis}.
  \newblock Cengage learning.

  \bibitem[\protect\citeauthoryear{Chen, Liu, and Zhang}{Chen
    et~al.}{2019}]{Chen2019Quantile}
  Chen, X., W.~Liu, and Y.~Zhang. 2019.
  \newblock Quantile regression under memory constraint.
  \newblock {\em Ann. Statist.\/}~{\em 47\/}(6): 3244--3273.
  \newblock \doi{10.1214/18-AOS1777} .

  \bibitem[\protect\citeauthoryear{Das, Pratama, Savitri, and Zhang}{Das
    et~al.}{2019}]{Das2019MUSE}
  Das, M., M.~Pratama, S.~Savitri, and J.~Zhang 2019.
  \newblock Muse-rnn: A multilayer self-evolving recurrent neural network for
    data stream classification.
  \newblock In {\em 2019 IEEE International Conference on Data Mining (ICDM)},
    pp.\  110--119.

  \bibitem[\protect\citeauthoryear{Fan and Gijbels}{Fan and
    Gijbels}{1992}]{Fan1992Variable}
  Fan, J. and I.~Gijbels. 1992.
  \newblock Variable bandwidth and local linear regression smoothers.
  \newblock {\em Ann. Statist.\/}~{\em 20\/}(4): 2008--2036.
  \newblock \doi{10.1214/aos/1176348900} .

  \bibitem[\protect\citeauthoryear{Fu, Zhao, and Zhou}{Fu
    et~al.}{2017}]{Fu2017Efficient}
  Fu, Y., W.~Zhao, and T.~Zhou. 2017.
  \newblock Efficient spectral sparse grid approximations for solving
    multi-dimensional forward backward {SDE}s.
  \newblock {\em Discrete Contin. Dyn. Syst. Ser. B\/}~{\em 22\/}(9): 3439--3458.
  \newblock \doi{10.3934/dcdsb.2017174} .

  \bibitem[\protect\citeauthoryear{Gautschi}{Gautschi}{2012}]{gautschi2011numerical}
  Gautschi, W. 2012.
  \newblock {\em Numerical analysis}.
  \newblock Springer Science \& Business Media, LLC, New York.

  \bibitem[\protect\citeauthoryear{Gutenbrunner and
    Jure\v{c}kov\'{a}}{Gutenbrunner and
    Jure\v{c}kov\'{a}}{1992}]{Gutenbrunner1992Regression}
  Gutenbrunner, C. and J.~Jure\v{c}kov\'{a}. 1992.
  \newblock Regression rank scores and regression quantiles.
  \newblock {\em Ann. Statist.\/}~{\em 20\/}(1): 305--330.
  \newblock \doi{10.1214/aos/1176348524} .

  \bibitem[\protect\citeauthoryear{Jiang, Qian, and Zhou}{Jiang
    et~al.}{2016}]{Jiang2016Single}
  Jiang, R., W.M. Qian, and Z.G. Zhou. 2016.
  \newblock Single-index composite quantile regression with heteroscedasticity
    and general error distributions.
  \newblock {\em Statist. Papers\/}~{\em 57\/}(1): 185--203.
  \newblock \doi{10.1007/s00362-014-0646-y} .

  \bibitem[\protect\citeauthoryear{Kai, Li, and Zou}{Kai et~al.}{2010}]{Kai2010}
  Kai, B., R.~Li, and H.~Zou. 2010.
  \newblock Local composite quantile regression smoothing: an efficient and safe
    alternative to local polynomial regression.
  \newblock {\em J. R. Stat. Soc. Ser. B Stat. Methodol.\/}~{\em 72\/}(1):
    49--69.
  \newblock \doi{10.1111/j.1467-9868.2009.00725.x} .

  \bibitem[\protect\citeauthoryear{Koenker}{Koenker}{2005}]{Koenker2005}
  Koenker, R. 2005.
  \newblock {\em Quantile regression}, Volume~38 of {\em Econometric Society
    Monographs}.
  \newblock Cambridge University Press, Cambridge.

  \bibitem[\protect\citeauthoryear{Koenker and Portnoy}{Koenker and
    Portnoy}{1987}]{Koenker1987Lestimation}
  Koenker, R. and S.~Portnoy. 1987.
  \newblock {$L$}-estimation for linear models.
  \newblock {\em J. Amer. Statist. Assoc.\/}~{\em 82\/}(399): 851--857 .

  \bibitem[\protect\citeauthoryear{Koenker and Zhao}{Koenker and
    Zhao}{1994}]{Koenker1994Lesti}
  Koenker, R. and Q.S. Zhao. 1994.
  \newblock {$L$}-estimation for linear heteroscedastic models.
  \newblock {\em J. Nonparametr. Statist.\/}~{\em 3\/}(3-4): 223--235.
  \newblock \doi{10.1080/10485259408832584} .

  \bibitem[\protect\citeauthoryear{Lin, Li, Wang, and Zhu}{Lin
    et~al.}{2019}]{LinComposite2019}
  Lin, L., F.~Li, K.~Wang, and L.~Zhu. 2019.
  \newblock Composite estimation: an asymptotically weighted least squares
    approach.
  \newblock {\em Statist. Sinica\/}~{\em 29\/}(3): 1367--1393 .

  \bibitem[\protect\citeauthoryear{Lin, Li, and Lu}{Lin
    et~al.}{2020}]{lin2020unified}
  Lin, L., W.~Li, and J.~Lu. 2020.
  \newblock Unified rules of renewable weighted sums for various online updating
    estimations.
  \newblock {\href{https://arxiv.org/abs/2008.08824}{{arXiv:2008.08824}}}
    {[stat.ME]}.

  \bibitem[\protect\citeauthoryear{Luo and Song}{Luo and
    Song}{2020}]{Luo2020Renewable}
  Luo, L. and P.X.K. Song. 2020.
  \newblock Renewable estimation and incremental inference in generalized linear
    models with streaming data sets.
  \newblock {\em J. R. Stat. Soc. Ser. B. Stat. Methodol.\/}~{\em 82\/}(1):
    69--97 .

  \bibitem[\protect\citeauthoryear{Moroshko, Vaits, and Crammer}{Moroshko
    et~al.}{2015}]{Moroshko2015Second}
  Moroshko, E., N.~Vaits, and K.~Crammer. 2015.
  \newblock Second-order non-stationary online learning for regression.
  \newblock {\em J. Mach. Learn. Res.\/}~16: 1481--1517 .

  \bibitem[\protect\citeauthoryear{Nion and Sidiropoulos}{Nion and
    Sidiropoulos}{2009}]{Nion2009Adaptive}
  Nion, D. and N.D. Sidiropoulos. 2009.
  \newblock Adaptive algorithms to track the parafac decomposition of a
    third-order tensor.
  \newblock {\em IEEE Transactions on Signal Processing\/}~{\em 57\/}(6): 2299
    – 2310.
  \newblock \doi{10.1109/TSP.2009.2016885} .

  \bibitem[\protect\citeauthoryear{Portnoy and Koenker}{Portnoy and
    Koenker}{1989}]{Portnoy1989Adaptive}
  Portnoy, S. and R.~Koenker. 1989.
  \newblock Adaptive {$L$}-estimation for linear models.
  \newblock {\em Ann. Statist.\/}~{\em 17\/}(1): 362--381.
  \newblock \doi{10.1214/aos/1176347022} .

  \bibitem[\protect\citeauthoryear{Pratama, Za'in, Ashfahani, Ong, and
    Ding}{Pratama et~al.}{2019}]{Pratama2019Automatic}
  Pratama, M., C.~Za'in, A.~Ashfahani, Y.S. Ong, and W.~Ding 2019.
  \newblock Automatic construction of multi-layer perceptron network from
    streaming examples.
  \newblock In {\em Proceedings of the 28th ACM International Conference on
    Information and Knowledge Management}, CIKM '19, New York, NY, USA, pp.\
    1171–1180. Association for Computing Machinery.

  \bibitem[\protect\citeauthoryear{Robbins and Monro}{Robbins and
    Monro}{1951}]{Herbert1951Stochastic}
  Robbins, H. and S.~Monro. 1951.
  \newblock {A Stochastic Approximation Method}.
  \newblock {\em The Annals of Mathematical Statistics\/}~{\em 22\/}(3): 400 --
    407.
  \newblock \doi{10.1214/aoms/1177729586} .

  \bibitem[\protect\citeauthoryear{Sauer}{Sauer}{2011}]{sauer2011numerical}
  Sauer, T. 2011.
  \newblock {\em Numerical analysis}.
  \newblock Addison-Wesley Publishing Company.

  \bibitem[\protect\citeauthoryear{Schifano, Wu, Wang, Yan, and Chen}{Schifano
    et~al.}{2016}]{Schifano2016}
  Schifano, E.D., J.~Wu, C.~Wang, J.~Yan, and M.H. Chen. 2016.
  \newblock Online updating of statistical inference in the big data setting.
  \newblock {\em Technometrics\/}~{\em 58\/}(3): 393--403.
  \newblock \doi{10.1080/00401706.2016.1142900} .

  \bibitem[\protect\citeauthoryear{Serfling}{Serfling}{1980}]{Serfling1980Approximation}
  Serfling, R.J. 1980.
  \newblock {\em Approximation theorems of mathematical statistics}.
  \newblock Wiley Series in Probability and Mathematical Statistics. John Wiley
    \& Sons, Inc., New York.

  \bibitem[\protect\citeauthoryear{Stigler}{Stigler}{1977}]{Stigler1977Do}
  Stigler, S.M. 1977.
  \newblock Do robust estimators work with real data?
  \newblock {\em Ann. Statist.\/}~{\em 5\/}(6): 1055--1098 .

  \bibitem[\protect\citeauthoryear{Sun, Gai, and Lin}{Sun
    et~al.}{2013}]{Sun2013Weighted}
  Sun, J., Y.~Gai, and L.~Lin. 2013.
  \newblock Weighted local linear composite quantile estimation for the case of
    general error distributions.
  \newblock {\em J. Statist. Plann. Inference\/}~{\em 143\/}(6): 1049--1063.
  \newblock \doi{10.1016/j.jspi.2013.01.002} .

  \bibitem[\protect\citeauthoryear{Toulis, Rennie, and Airoldi}{Toulis
    et~al.}{2014}]{Toulis2014Statistical}
  Toulis, P., J.~Rennie, and E.~Airoldi. 2014.
  \newblock Statistical analysis of stochastic gradient methods for generalized
    linear models.
  \newblock {\em International Conference on Machine Learning\/}~{\em 32\/}(1):
    667 – 675 .

  \bibitem[\protect\citeauthoryear{Wang, Wang, and Li}{Wang
    et~al.}{2022}]{Wang2022Renewable}
  Wang, K., H.~Wang, and S.~Li. 2022.
  \newblock Renewable quantile regression for streaming datasets.
  \newblock {\em Knowledge-Based Systems\/}~235: 107675.
  \newblock \doi{https://doi.org/10.1016/j.knosys.2021.107675} .

  \bibitem[\protect\citeauthoryear{Zhao, Chen, and Peng}{Zhao
    et~al.}{2006}]{Zhao2006new}
  Zhao, W., L.~Chen, and S.~Peng. 2006.
  \newblock A new kind of accurate numerical method for backward stochastic
    differential equations.
  \newblock {\em SIAM J. Sci. Comput.\/}~{\em 28\/}(4): 1563--1581.
  \newblock \doi{10.1137/05063341X} .

  \bibitem[\protect\citeauthoryear{Zhao, Fu, and Zhou}{Zhao
    et~al.}{2014}]{Zhao2014New}
  Zhao, W., Y.~Fu, and T.~Zhou. 2014.
  \newblock New kinds of high-order multistep schemes for coupled forward
    backward stochastic differential equations.
  \newblock {\em SIAM J. Sci. Comput.\/}~{\em 36\/}(4): A1731--A1751.
  \newblock \doi{10.1137/130941274} .

  \bibitem[\protect\citeauthoryear{Zou and Yuan}{Zou and
    Yuan}{2008}]{Zou2008Composite}
  Zou, H. and M.~Yuan. 2008.
  \newblock Composite quantile regression and the oracle model selection theory.
  \newblock {\em Ann. Statist.\/}~{\em 36\/}(3): 1108--1126.
  \newblock \doi{10.1214/07-AOS507} .

  \end{thebibliography}
\end{document}